\documentclass[twocolumn]{aa}
\usepackage{graphicx}
\usepackage{txfonts}

\usepackage{times}
\usepackage{graphics}
\usepackage{epsf}
\def\plotfiddle#1#2#3#4#5#6#7{\centering \leavevmode
\vbox to#2{\rule{0pt}{#2}}
\includegraphics{#1}}

\begin{document}
\title{Properties of dust in the high-latitude translucent cloud L1780}
\subtitle{I: Spatially distinct dust populations and increased dust emissivity from ISO observations*}
\thanks{Based on observations with ISO, an ESA project with instruments funded by ESA Member States (especially the PI countries: France, Germany, the Netherlands and the United Kingdom) and with the participation of ISAS and NASA}

   \author{M. Ridderstad
          \inst{1}
          \and
	  M. Juvela\inst{1}
	  \and
	  K. Lehtinen\inst{1}
	  \and
	  D. Lemke\inst{2}
	  \and
	  T. Liljestr\"om\inst{1}
          }

   \offprints{M. Ridderstad \email{marianna@astro.helsinki.fi}}

   \institute{Helsinki University Observatory, T\"ahtitorninm\"aki,
              P.O. Box 14, FIN-00014 University of Helsinki, Finland\\
         \and
              Max-Planck-Institut f\"ur Astronomie, K\"onigstuhl 17, 
	      D-69117, Germany\\
             }

\date{Received ; accepted}

\abstract{We have analyzed the properties of dust in the high galactic
latitude translucent cloud Lynds 1780 using ISOPHOT maps at 100$\mu$m and
200$\mu$m and raster scans at 60$\mu$m, 80$\mu$m, 100$\mu$m, 120$\mu$m,
150$\mu$m and 200$\mu$m. In far-infrared (FIR) emission, the cloud has a
single core that coincides with the maxima of visual extinction and 200$\mu$m
optical depth. At the resolution of 3.0$\arcmin$, the maximum visual
extinction is 4.0 mag. At the cloud core, the minimum temperature and the
maximum 200$\mu$m optical depth are 14.9$\pm0.4$ K and $2.0\pm0.2\times
10^{-3}$, respectively, at the resolution of 1.5$\arcmin$. The cloud mass is
estimated to be 18$M_{\sun}$. The FIR observations, 
combined with IRAS observations, suggest the presence of
different, spatially distinct dust grain populations in the cloud: the FIR
core region is the realm of the "classical" large grains, whereas the very
small grains and the PAHs have separate maxima on the Eastern side of the cold
core, towards the "tail" of this cometary-shaped cloud. The color 
ratios indicate an overabundance of PAHs and VSGs in L1780. 
Our FIR observations combined with the optical extinction data
indicate an increase of the emissivity of the big grain dust component in the
cold core, suggesting grain coagulation or some other change in the properties of the
large grains. Based on our observations, we also address the question, to what
extent the 80$\mu$m emission and even the 100$\mu$m and the 120$\mu$m emission
contain a contribution from the small-grain component.  
   
}

\maketitle
%

\section{Introduction} 

Recent ISM dust models (e.g. D\'esert, Boulanger \& Puget 1990) 
typically have three
dust components for which a power law size distribution is assumed: the big
"classical'' silicate or carbonaceous dust grains with maximum emission at
about 100-200$\mu$m; very small grains (VSGs), which mainly
emit at $\lambda=20-60\mu$m; and polycyclic aromatic hydrocarbons
(PAHs) which have
emission features in the wavelength range 3-20$\mu$m and a possible continuum
contribution. It has been known since the IRAS observations that both the
small and the big interstellar grains contribute to the heating and cooling of
the interstellar medium (ISM), and it is believed that also PAHs contribute
significantly to the heating process (Bakes \& Tielens 1994). In a typical
translucent cloud, such as L1780, heating takes place mainly through
photoelectrons that are created when energetic photons from the interstellar
radiation field (ISRF) hit dust grains. 
When heated by the solar neighborhood ISRF, 
large grains are able to remain at an equilibrium
temperature, emitting mostly at far-infrared (FIR) wavelengths.
On the other hand, when small grains or PAHs absorb an
energetic UV photon they are temporarily heated to higher temperatures, 
which causes their temperature to fluctuate.
While small grains, which radiate mostly at shorter 
wavelengths (below 100$\mu$m), 
play a significant part in both the heating and cooling of
clouds, their relationship to the large grains, their relative abundance, as
well as their internal properties, are still incompletely understood.

COBE and ISO satellites extended the infrared observational range to longer
wavelengths, enabling the measurement of the large-grain temperatures and
emissivities. Observations show that the spectral energy distribution of the
interstellar dust varies from the diffuse medium to molecular clouds and also
from cloud to cloud.  This probably implies changes in the dust composition,
and especially variations in the abundance of VSGs and PAHs (Boulanger et al.
1990, Lagache et al. 1998, Verter et al. 2000).  Variations in the abundances
of grain populations within a single cloud have also been suggested 
(Bernard et al. 1992, 1993; Rawlings et al. 2005). Moreover,
the observed very cold temperatures and changes in the emissivity of the large
grains also suggest that the properties of the large grains change in regions
of high dust column density in the interiors of molecular clouds (Bernard et
al. 1999, Stepnik et al. 2003). In this paper, and in a subsequent paper,
where a radiative transfer model of L1780 will be presented (Ridderstad et
al., in preparation), we address some of the above questions through
observations obtained towards the translucent cloud Lynds 1780.

Lynds 1780 (hereafter L1780; also listed as MBM 33 in 
Magnani, Blitz \& Mundy (1985) is
a high galactic latitude ($l=359.0\degr$, $b=36.7\degr$) cloud with the size
of $\sim40\arcmin\times30\arcmin$. The brighter central part of the cloud was 
listed as a separate source, L1778, in Lynds' (1962) catalogue.  L1780 was
classified as a cometary globule associated with Loop I by T\'oth et al.
(1995). It is located in the dark cloud complex that includes the clouds L134,
L169, and L183 (L134N), and is likely to be physically associated with it
(Clark \& Johnson 1981). We refer to these clouds, including L1780, as the
L134 complex. The distance to the L134 complex was determined to be 110$\pm
10$ pc by Str\"omgren photometry (Franco 1989) and 140$\pm 20$ pc by
Vilnius photometry (Cernis \& Straizys 1992). The NaI measurements by
Lallement et al. (2003) indicate a distance of $\sim 100$ pc.  In this paper,
we adopt the distance of 110 pc. 

The L134 complex is the 
northernmost extension of the Scorpius-Ophiuchus star-forming region 
and is located at the border of the Local Bubble 
(Kuntz, Snowden \& Verter 1997, Lallement et al. 2003). In addition to L1780, 
also other clouds in this complex have displaced, cometary-like cores. 
In L183, very early phases of star formation are indicated 
(Ward-Thompson et al. 1994, 2000; Lehtinen et al. 2003). 
No star formation has been observed in L1780, although young
H$\alpha$ emission line stars with ages a few times $10^6$ years 
have been observed around L1780 (Martin \& Kun 1996).
One of these H$\alpha$ stars has the same radial velocity as 
L1780 (Martin \& Kun 1996), and they may be related to an old SNR 
expansion shell (T\'oth et al. 1995, Martin \& Kun 1996).

Observations of HI in L1780 have shown that, on its Southern half, the cloud has
excess emission, which is probably caused by the UV-radiation of the bright 
OB stars in the direction of the galactic plane (Mattila \& Sandell 1979).
T\'oth et al. (1995) discussed in detail the virial equilibrium conditions and
the evolutionary history of L1780, and suggested that the cometary structure
of the cloud is produced by the shock fronts of supernovae and stellar winds
from the Sco-Cen association of OB stars. They also found that the $^{13}$CO
core, which coincides with the IRAS 100$\mu$m emission core, is in 
virial equilibrium.

The maximum optical extinction $A_B$ in the cloud has been estimated to be 4
mag by Mattila (1986). He also found a good correlation between CH column
density and optical extinction in L1780.
The HI and CO distributions in L1780 (Mattila \& Sandell 1979, T\'oth et al.
1995) were correlated with different IRAS band distributions, which suggested
that there are different grain populations present in L1780, and their
distributions are distinct from each other and related to the local physical
conditions. 
Laureijs (1989) studied L1780 in the optical and infrared, and
suggested that the optical excess red emission observed in the cloud by
Mattila (1979) is analogous to that observed in the Red Rectangle 
(Schmidt, Cohen \& Margon al. 1980). 
If this interpretation is correct, L1780 is so far the only
translucent cloud in which this Extended Red Emission (ERE) has been observed.

In this paper, we study the 60$\mu$m-200$\mu$m emission of L1780 using ISO
observations. We also compare these data with the IRAS observations at
12$\mu$m and 25$\mu$m to investigate the PAH abundance within L1780, as the
radiation at 12$\mu$m is generally believed to be due to PAHs, and the
25$\mu$m emission is mainly from VSGs (D\'esert, Boulanger \& Puget 1990).
While the radiation at 60$\mu$m is due to VSGs, the radiation at
$\sim$100$\mu$m-200$\mu$m can be described as a modified blackbody emission
from the 'classical big' dust grains, which are in equilibrium with the
surrounding radiation field.  It is not entirely clear, to what extent the
100$\mu$m emission contains a small-grain contribution; we compare our ISO
raster scan observations to the current ISM dust model 
to address this question.
However, since the small-grain contribution at 100$\mu$m is small, the
100$\mu$m-200$\mu$m maps and raster scans can be used to effectively trace the
domain of the large grains, which are responsible for most of the interstellar
extinction at optical wavelengths and make up most of the dust mass.  We use
the 100$\mu$m and 200$\mu$m ISO observations to derive estimates on the dust
temperature, column density and mass. From the 60$\mu$m, 80$\mu$m, 100$\mu$m,
120$\mu$m, 150$\mu$m and 200$\mu$m ISO raster scans, the spectral energy
distributions (SEDs) and temperatures at selected positions of the cloud are
obtained. 2MASS data are used to derive an optical extinction map. The
locations of the maxima and the spatial distributions of the emission at 
different wavelengths are compared, and the distributions of the three dust
components (PAHs, VSGs and the large grains) are estimated. We also compare
our results on L1780 with the recent observations suggesting that the 
FIR emissivity of dust increases where low dust temperatures
are observed (Cambr\'esy et al. 2001, del Burgo et al. 2003, Stepnik et al.
2003, Kramer et al. 2003). 


\begin{figure*}
\begin{minipage}{12cm}
\plotfiddle{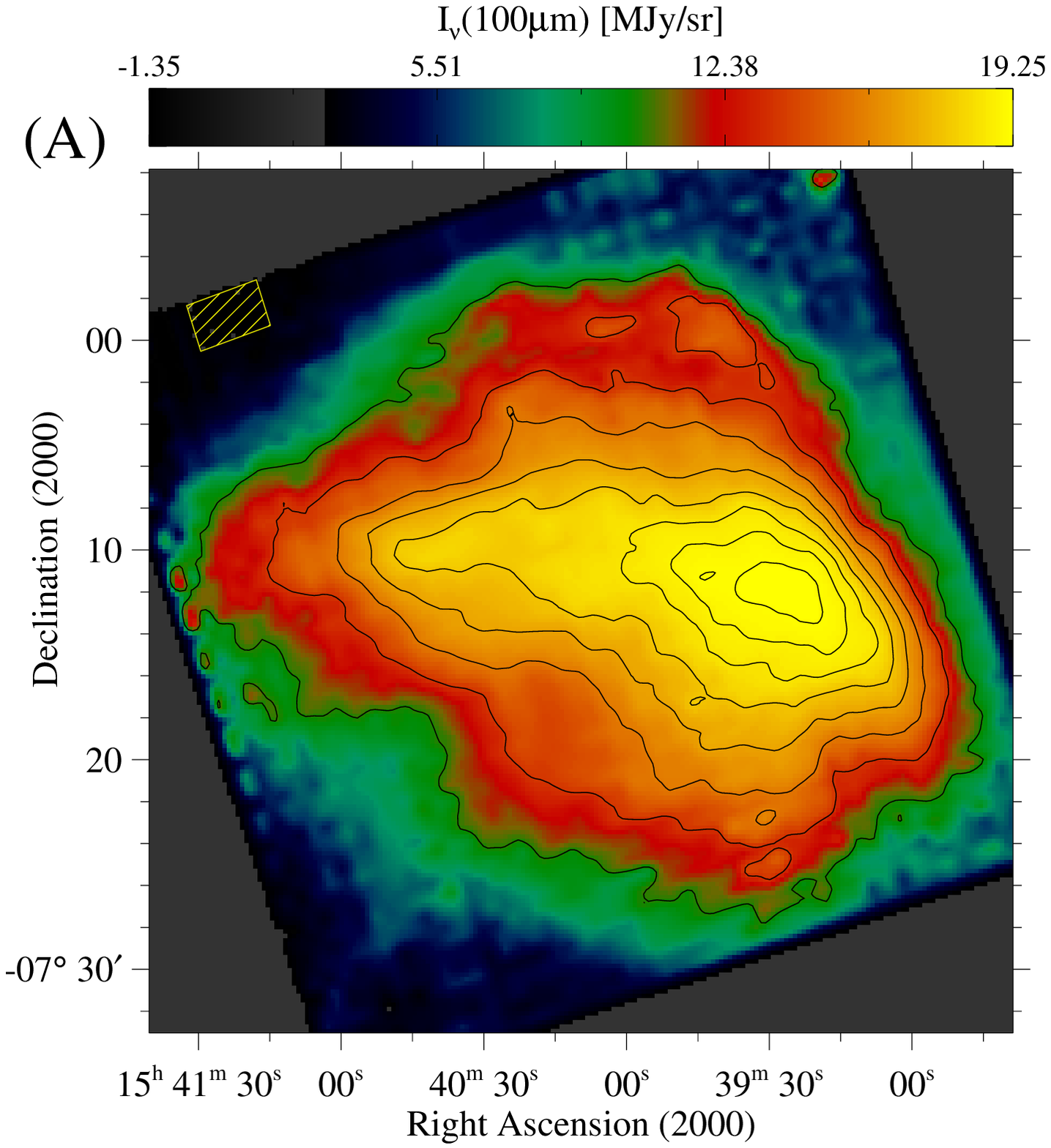}{6cm}{0}{44}{44}{-130}{-14}
\plotfiddle{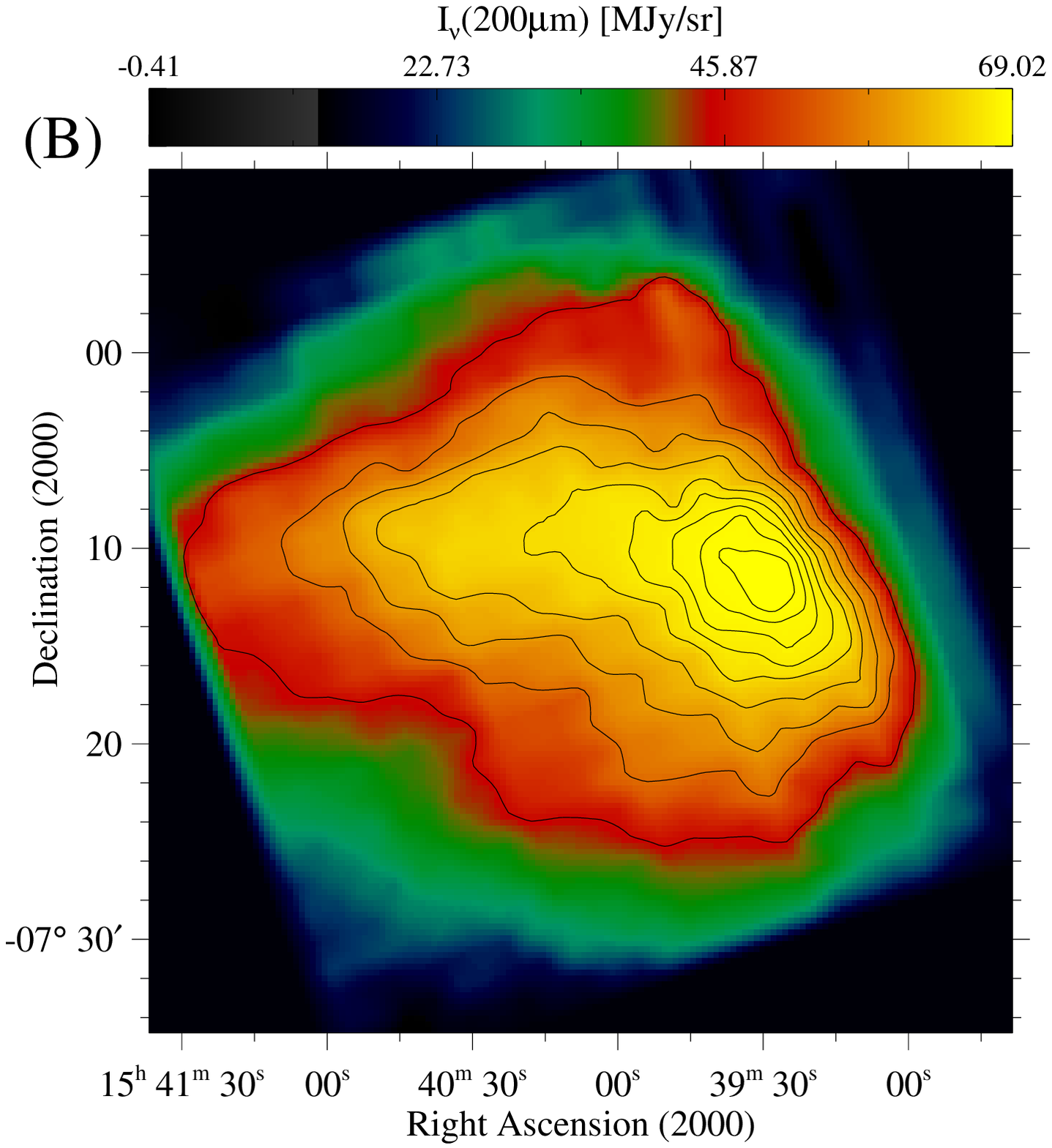}{0cm}{0}{44}{44}{+70}{+8}
\plotfiddle{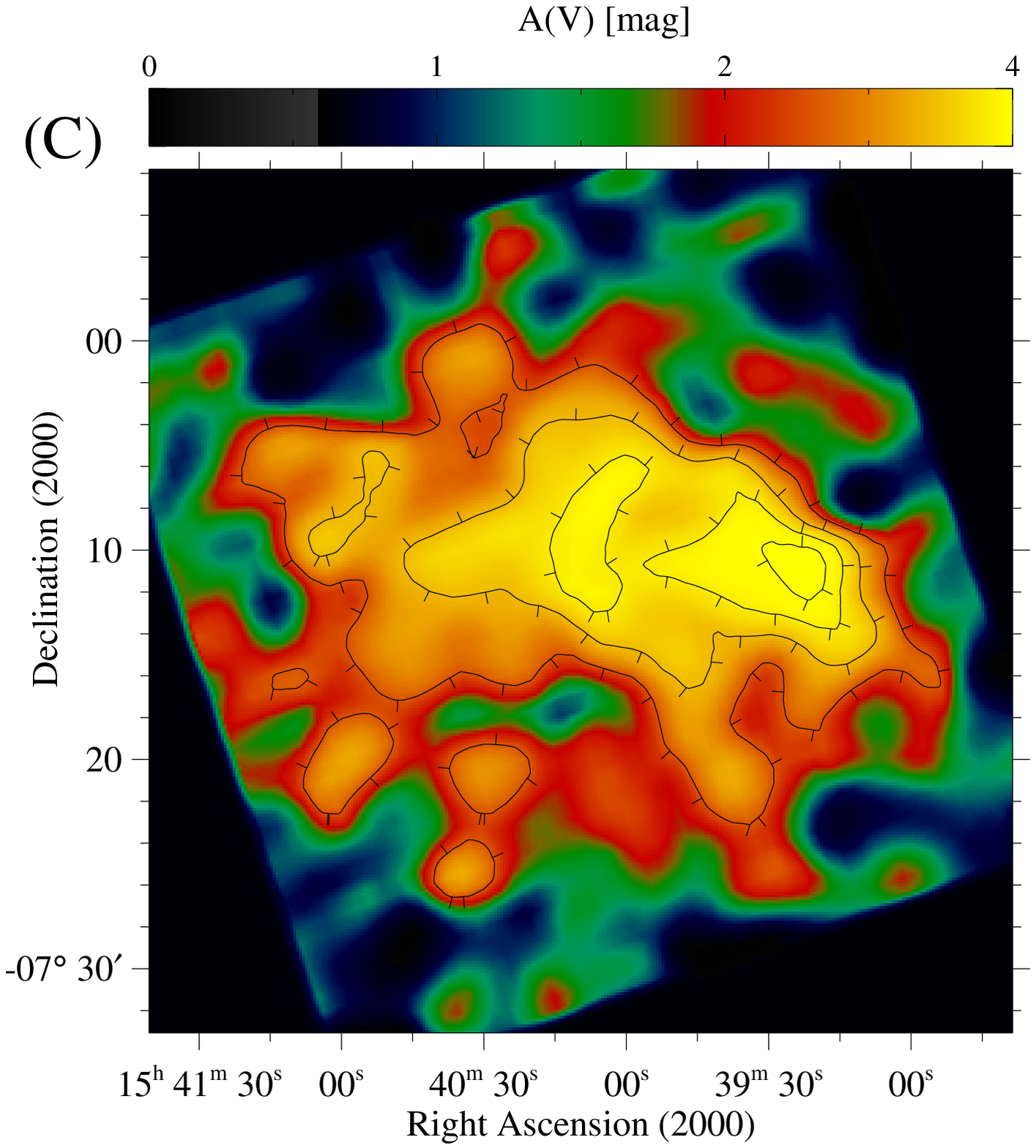}{6cm}{0}{44}{44}{-130}{-14}
\plotfiddle{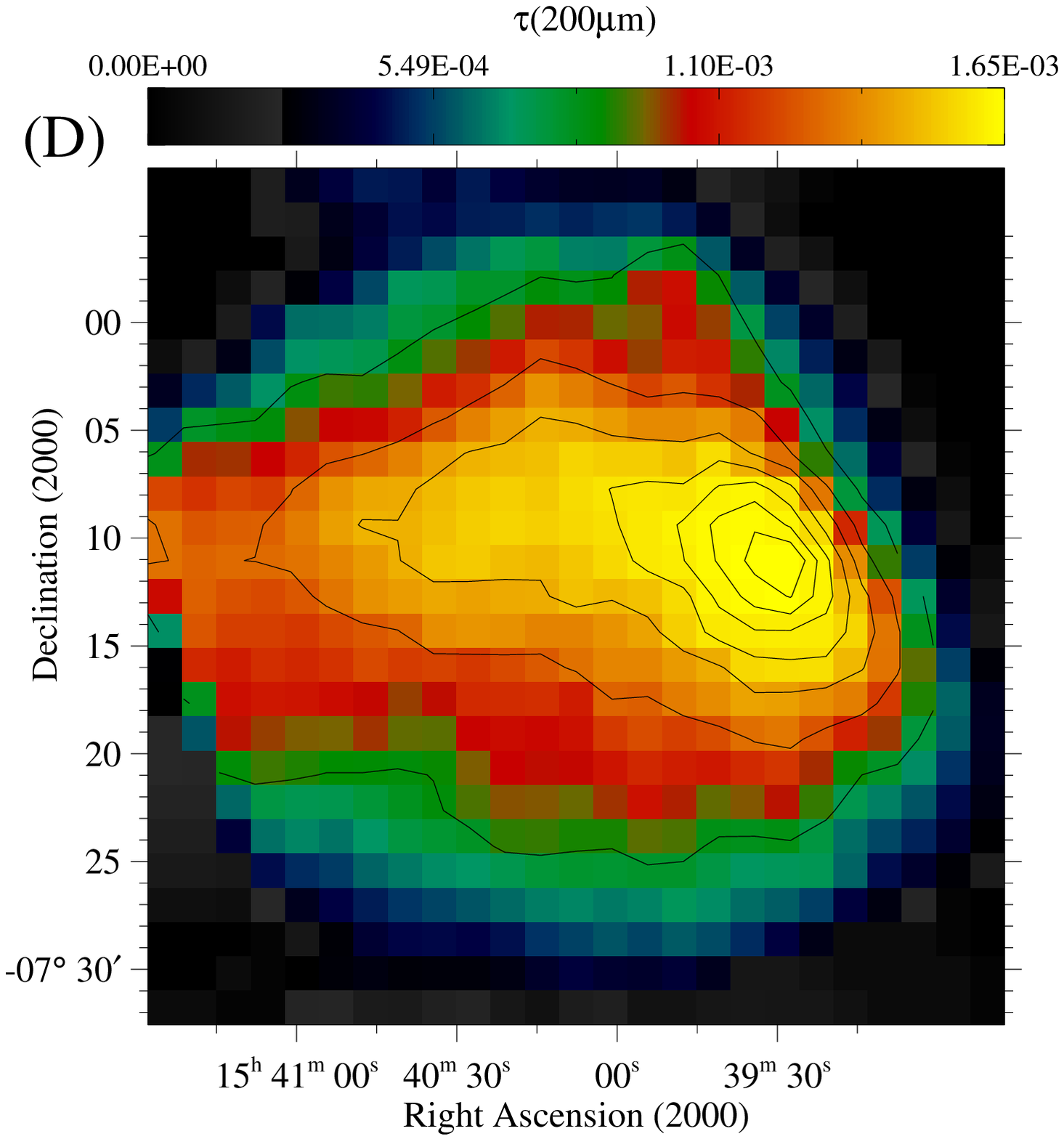}{0cm}{0}{44}{44}{+70}{+8}
\plotfiddle{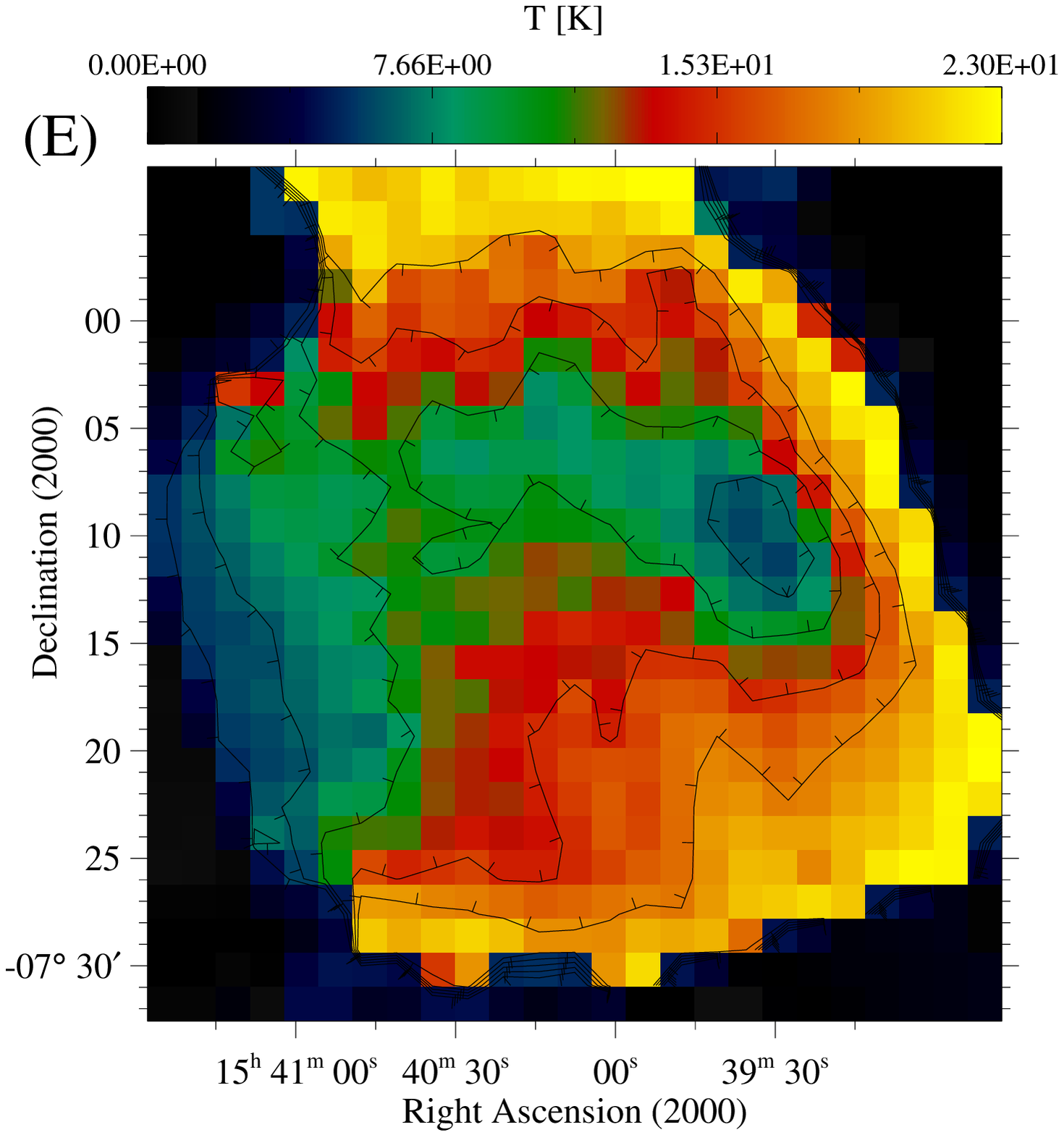}{6cm}{0}{44}{44}{-130}{-14}
\plotfiddle{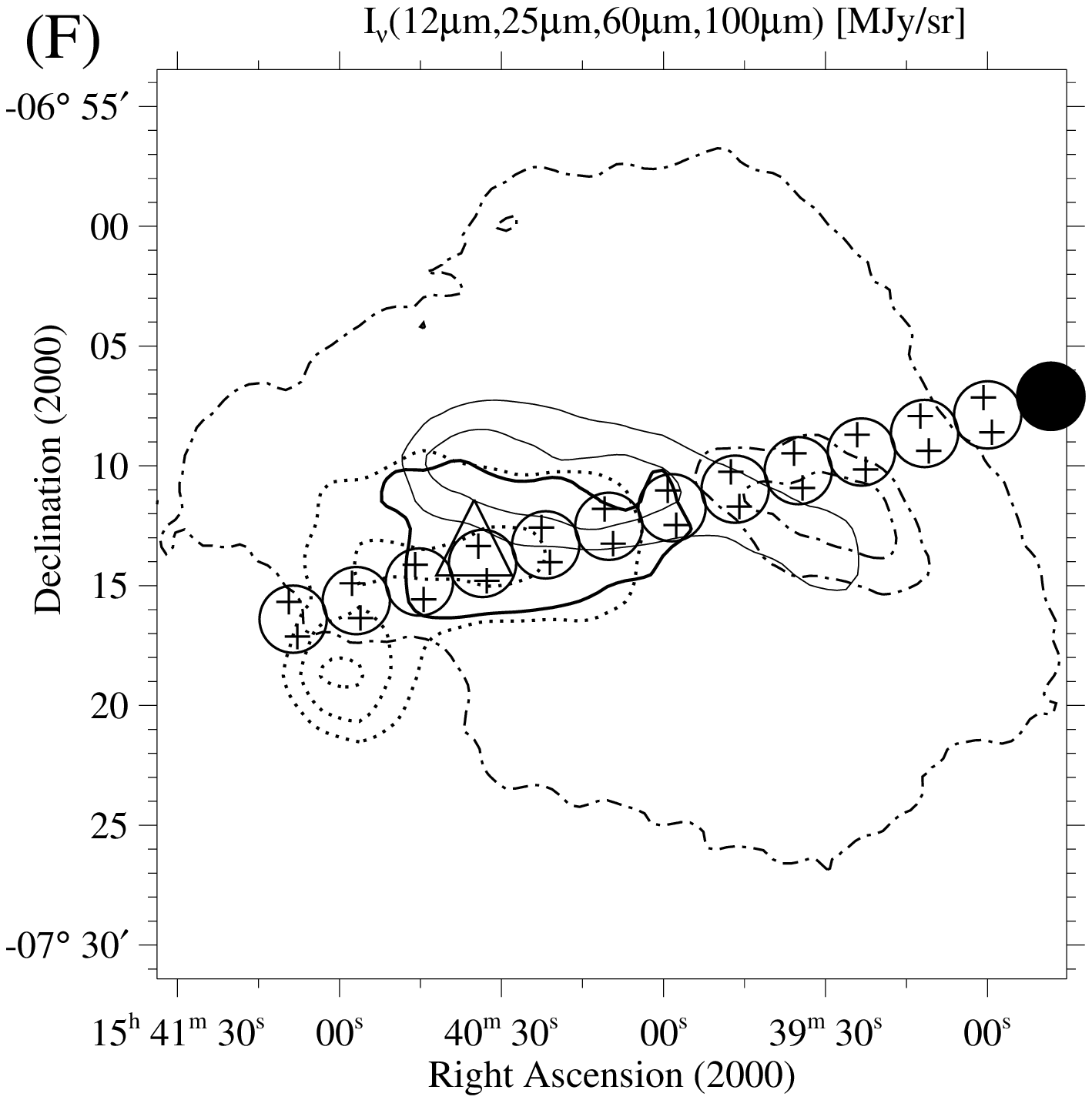}{0cm}{0}{44}{44}{+70}{+8}
\end{minipage}
\caption[]{
(a) The 100$\mu$m ISO surface brightness map of L1780 at 43.5$\arcsec$ resolution. The contour values range from 4.0 to 18.0 ~MJy~sr$^{-1}$ in steps of 2.0~MJy~sr$^{-1}$. The area used for the background sky value determination and subtraction is marked with a rectangle. 
(b) The 200$\mu$m ISO surface brightness map of L1780 at the resolution of 1.5$\arcmin$, with contours from 13~MJy~sr$^{-1}$ to 61~MJy~sr$^{-1}$ in steps of 6~MJy~sr$^{-1}$.
(c) The visual extinction map of L1780 at 3.0$\arcmin$ resolution, with contours from 0.8 mag to 4.0 mag in steps of 0.8 mag, i.e. one step corresponds to about $2\sigma$. The tickmarks point towards decreasing $A_V$.
(d) The 200$\mu$m optical depth map of L1780 at 1.5$\arcmin$ resolution, with contours from 2.0$\times 10^{-4}$ to 16.0$\times 10^{-4}$ in steps of 2.0$\times 10^{-4}$.
(e) The temperature map of L1780 at the resolution of 1.5$\arcmin$, with contours from 15.2 K to 16.8 K in steps of 0.5 K. The tickmarks point towards decreasing $T$.
(f) The 100$\mu$m ISO map showing the positions of the IRAS 12$\mu$m (dotted line), IRAS 25$\mu$m (thick solid line), IRAS 60$\mu$m (thin solid line), ISO 100$\mu$m (dash-dot line) and ISOCAM 6.7$\mu$m (large triangle) maxima (Miville-Desch\^enes 2002), 
and the positions of the detector center of an ISO raster scan (marked with crosses).
The cirles show the positions used in calculating the averaged values of the raster scans (see the text). The black circle indicates the sky position used for background subtraction for the raster scans. 
On the SE side of the L1780, the IRAS 12$\mu$m contours show a point source, IRAS 15383-0709, which is probably not associated with the cloud.
}
\label{fig:multiplot}
\end{figure*}

\section{Observations and data reduction}

\subsection{FIR observations}
The observations were made with the ISOPHOT instrument aboard the Infrared 
Space Observatory (ISO) (Kessler, Steinz \& Anderegg 1996) 
satellite, using the C100 and C200 detectors (Lemke et al. 1996)
with the observing template PHT22 in raster mode. 
The data analysis was done with PIA (ISOPHOT 
Interactive Analysis) V10.0 (Gabriel et al. 1997). At the 
first processing level, the detector ramps were corrected for non-linearity 
in the detector response, glitches in ramps were removed using the 
two-threshold glitch recognition method, and the ramps were fitted 
with 1st order polynomials. At subsequent levels the signals were 
deglitched, reset interval correction was applied, signals were linearized 
for the dependence of the detector response on illumination, and orbital 
position-dependent dark currents were subtracted.

\begin{table*}
\caption[]{The parameters of the individual maps and raster scans. $\lambda_{ref}$is the PHT reference wavelength,  $\lambda_c$ is the central wavelength, $\Delta\lambda$ is the width, and TDT is the Target Dedicated Time number of the observation.}
\begin{center}
\begin{tabular}{rrrrrrrr}
\hline\noalign{\smallskip}
Filter  &  $\lambda_{ref}$ & $\lambda_c$  &  $\Delta\lambda$ &  TDT  &  Raster steps &  Map size  &  Camera                           \\
        & $\mu$m  &  $\mu$m & $\mu$m &  &  &  & \\
\noalign{\smallskip}
\hline\noalign{\smallskip}
C100 & 100 & 102.6 & 47.1 & 43100630 & 18$\times$24 & $40\arcmin\times 36\arcmin$& C100\\
C200 & 200 & 202.1 & 56.9 & 43199629 & 13$\times$13 & $39\arcmin\times 39\arcmin$& C200\\
C60 & 60 & 61.8 & 24.6 & 43100206 & 2$\times$16 & $47\arcmin\times 3.7\arcmin$ & C100\\
C70 & 80 & 80.7 & 48.4 & 43100207 & 2$\times$16 & $47\arcmin\times 3.7\arcmin$ & C100\\
C100 & 100 & 102.6 & 47.1 & 43100208 & 2$\times$16 & $47\arcmin\times 3.7\arcmin$ & C100\\
C120 & 120 & 118.7 & 49.5 & 43100209 & 2$\times$13 & $39\arcmin\times 4.5\arcmin$ & C200\\
C135 & 150 & 155.1 & 81.2 & 43100212 & 2$\times$13 & $39\arcmin\times 4.5\arcmin$ & C200\\
C200 & 200 & 202.1 & 56.9 & 43100210 & 2$\times$13 & $39\arcmin\times 4.5\arcmin$ & C200\\
\noalign{\smallskip}
\hline
\end{tabular}
\end{center}
\label{tab:obsdata}
\end{table*}

Table~\ref{tab:obsdata} shows the parameters of the individual observations.
All the maps were calibrated using the 
FCS (Fine Calibration Source) measurements bracketing the actual 
measurements. The sizes of the map pixels of the C100 maps and 
the C200 maps are 43.5$\arcsec$ and 89$\arcsec$, respectively. A typical 
statistical uncertainty is 0.2~MJy~sr$^{-1}$ for a map pixel of 
the large 100$\mu$m map and 1.2~MJy~sr$^{-1}$ for a 200$\mu$m map pixel.
For the raster scans, the average uncertainties are 0.4, 0.4, 0.3, 1.5, 0.1 
and 0.2~MJy~sr$^{-1}$ for
a 60$\mu$m, 80$\mu$m, 100$\mu$m, 120$\mu$m, 150$\mu$m and 200$\mu$m map pixel,
respectively. The uncertainty in the absolute calibration is 
$\sim$25\% for the C100 and $\sim$20\% for the C200 data (Klaas et al. 2000).

For the flat-field correction of the large C100 and C200 maps
a statistical method was applied: 
the pixel values were correlated against the reference pixel at each raster
position. Rather than comparing the reference pixel only with
pixels belonging to the same raster position, a mean of two
pixels located symmetrically around the reference pixel was taken. This
reduces the scatter in the pixel-to-pixel relation caused by surface
brightness gradients. For the flat-fielding of the 
ISO raster scans, a similar procedure was used.
The value of the reference pixel was calculated as a weighted average using 
a Gaussian that was placed at the position of the studied 
pixel and had a FWHM of 2.0$\arcmin$. 
For both the large ISO maps and the raster scans the 
reference pixels were chosen to be the pixels number 8 and 3 
in the C100 and C200 rasters, respectively. The data from pixel no 3 
of the ISO 60$\mu$m raster scan was of such bad quality that it was 
replaced by the average of the values of the surrounding raster pixels.

It is known that the primary intensity calibrators, the fine calibration 
source (FCS) measurements used to derive the responsivities of the detectors,
suffer from signal transients in the case of the C100 camera. 
Lehtinen et al. (2001) showed that even after these transients were corrected
using the PIA signal drift interface, there remained a difference of 20\%
between the 100$\mu$m ISO and DIRBE surface brigtnesses. Therefore, the C100
measurements on L1780 were compared with DIRBE.

IRAS data was first scaled to DIRBE scale using the relation 
$I_{\rm DIRBE}(100\mu{\rm m})=0.73\times I_{\rm IRAS}(100\mu{\rm m})$ that was
derived for a $6\degr$ diameter circular region around L1780.  A linear fit
ISO vs. scaled IRAS values gave $I_{\rm IRAS(DIRBE)}(100\mu{\rm m})=
0.61\times I_{\rm ISO}(100\mu{\rm m})$. In the comparison both DIRBE and ISO
observations were color corrected assuming a spectrum $B_{\nu}(T_{\rm dust}=17\,K) \nu^2$.
The larger 100$\mu$m map was scaled to the DIRBE surface brightness scale using
this relation, and the 100$\mu$m raster scan was adjusted to the same level
with the scaled raster map.

The 60$\mu$m raster scan values were similarly scaled to DIRBE via IRAS, 
using the relations $I_{\rm DIRBE}(60\mu{\rm m})= 0.69\times I_{\rm
IRAS}(60\mu{\rm m})$ and $I_{\rm IRAS(DIRBE)}(60\mu{\rm m})= 0.86\times I_{\rm
ISO}(60\mu{\rm m})$. No color correction was applied since at 60$\mu$m the
dust spectrum is rather close to a flat spectrum, $\nu I_{\nu}=$ constant.
We also compared the values calculated from the
zodiacal light model of Good (1994), to the corresponding empty-sky values of
the 60$\mu$m raster map (raster position within the circle marked in
Fig.~\ref{fig:multiplot}f) that were color corrected for a 270\,K black body spectrum.
This gave the relation $I_{model}(60\mu{\rm m})=
0.89\times I_{\rm ISO}(60\mu{\rm m})$.  We took the average of the two
scaling factors, obtaining $I_{true}(60\mu{\rm m})= 0.88\times I_{\rm
ISO}(60\mu{\rm m})$ as the final scaling relation.

The 80$\mu$m data was scaled by linearly interpolating 
the scaling coefficients for the 60$\mu$m and 100$\mu$m ISO data, 
which gave 0.71 as the scaling coefficient for 80$\mu$m. 

The IRAS 12$\mu$m and 25$\mu$m Infrared Sky Survey Atlas (ISSA) 
map data were re-scaled to DIRBE scale using the relations
$I_{\rm DIRBE}(12\mu{\rm m})= 1.06\times I_{\rm IRAS}(12\mu{\rm m})$ and 
$I_{\rm DIRBE}(25\mu{\rm m})= 1.01\times I_{\rm IRAS}(25\mu{\rm m})$, 
as given in the ISSA Explanatory Supplement (Wheelock et al. 1994).

The ISO 200$\mu$m map surface brightness values were also compared with 
DIRBE data. There are only 4 DIRBE pixels within the 200$\mu$m map area. The
DIRBE surface brightness values at 100$\mu$m, 140$\mu$m and 240$\mu$m were 
fitted with a modified blackbody function with $\lambda^{-2}$ emissivity
law in order to get interpolated values at 200$\mu$m. 
The ISO 200$\mu$m map was convolved with a scan-averaged DIRBE beam and 
the ISO surface brightness values were color corrected using 
the temperatures derived from fitting the DIRBE values. A linear 
fit forced to go through the origin gives 
$I_{\rm DIRBE}(200\mu{\rm m})= 0.98\times I_{\rm ISO}(200\mu{\rm m})$,
which is well within the estimated uncertainties of the ISO data. 
Since this fit is made based on four points only, which results 
in considerable uncertainty, and the C200 detector does not suffer
from such problems as described above for the C100 detector, 
we retain the FCS based calibration for the 200$\mu$m and other C200 ISO data 
(the 120$\mu$m and 150$\mu$m data).

The sky area used for subtracting the background values in the 100$\mu$m and
200$\mu$m maps is shown in Fig.~\ref{fig:multiplot}a.  The background sky
values for background subtraction for all the six raster scans and the IRAS
data used were determined at the same position outside the cloud
(Fig.~\ref{fig:multiplot}f); this way we obtained a common zero level for
comparison with the raster scans.  To compare the six raster scans with each
other, we calculated average values at 16 evenly-spaced positions along the
C100 scans, and at 13 positions along the (shorter) C200 scans (see
Fig.~\ref{fig:multiplot}f for these positions; three C100 positions in the
west fall outside the figure).  The averaging was done using a Gaussian weight
function with FWHM=4.5$\arcmin$. This resolution also corresponds to the
resolution of the ISSA maps.  

\begin{figure}
\centering 
\resizebox{\hsize}{!}{\includegraphics{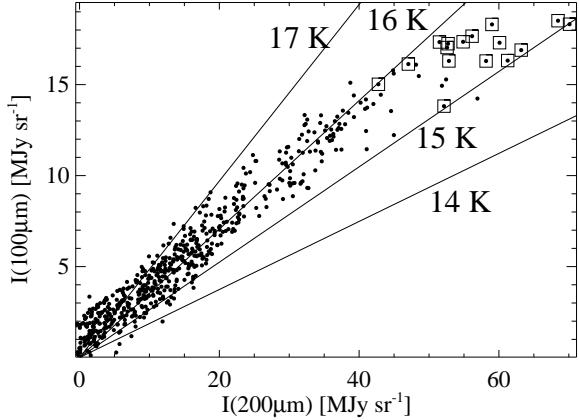}}
\caption{The 100$\mu$m vs. the 200$\mu$m ISO map values. The solid lines show the surface brightness ratios at the indicated temperatures, assuming an $\beta =2$ emissivity power law index. The points corresponding to the sky area of the ISO 100$\mu$m surface brightness maximum ($6.25\arcmin\times 4.0\arcmin$, corresponding to the highest level contour in Fig.~\ref{fig:multiplot}a) are marked with squares.}
\label{fig:c200vsc100}
\end{figure}

\subsection{Extinction data}
An extinction map of L1780 was derived using the $J$, $H$ and 
$K_{\rm_S}$ band magnitudes obtained from the 2MASS archive.
The extinction measurements are based on the near-infrared (NIR) 
color excesses of stars visible through the cloud. We applied the
optimized multi-band NICER technique of Lombardi \& Alves (2001), which is a
generalization of the traditional color excess method (using data on two
bands only), to derive the NIR color excesses. 

Stellar density of the detected stars at all three bands is fairly
constant over the map, with an average of about one
star per square arcminute. As the reference field, representing an area 
without significant extinction, we selected a 1$\degr$ diameter circular area
located close to L1780 and at the same galactic latitude. 
The coordinates    
of the centre of this area are R.A.(J2000)$=15^{\rm h}22^{\rm m}00^{\rm s}$, 
Dec.(J2000)$=-6\degr 05\arcmin 00\arcsec$. The colors of the stars in the
reference field have the mean values and standard deviations of
$\overline{(J-H)}_0=0.49\pm 0.19$ and $\overline{(H-K)}_0=0.12\pm 0.23$.
As the effective wavelengths of the 2MASS $J$, $H$ and 
$K_{\rm_S}$ bands we use 1.25$\mu$m, 1.65$\mu$m and 2.17$\mu$m, 
respectively (Kleinmann et al. 1994). For the ratios of visual 
extiction to color excess we used the values $A_V\slash E(J-H)=8.86$ 
and $A_V\slash E(H-K_{\rm_S})=15.98$, which
correspond to an extinction curve with $R_V=3.1$ (Mathis 1990).

The extinction value in each map
pixel was derived from the individual extinction values of the stars
by applying the sigma-clipping technique of Lombardi \& Alves
(2001).
In order to obtain a sufficient signal-to-noise ratio in the $A_V$ map,
the individual extinction values were averaged using a Gaussian with
FWHM=3.0$\arcmin$.

The extinction map has two error sources, the variance of the intrinsic 
colors $(J-H)_0$ and $(H-K)_0$, and the variance of the observed magnitudes 
of the field stars. The former dominates with a 1$\sigma$ 
error of $\sim 0.42 {\rm mag}$ per pixel, while the latter gives a 1$\sigma$
error of $\sim 0.23 {\rm mag}$ per pixel, 
resulting in a typical total error of 0.47 mag in $A_V$.

The map of visual extinction is shown in Fig.~\ref{fig:multiplot}c. At the
resolution of 3.0$\arcmin$ the value of maximum extinction is 4.0 mag. 

When the extinction data is compared with the 100$\mu$m and 200$\mu$m surface
brightness, temperature and optical depth maps, the latter are convolved to
this resolution.

\begin{figure}
\plotfiddle{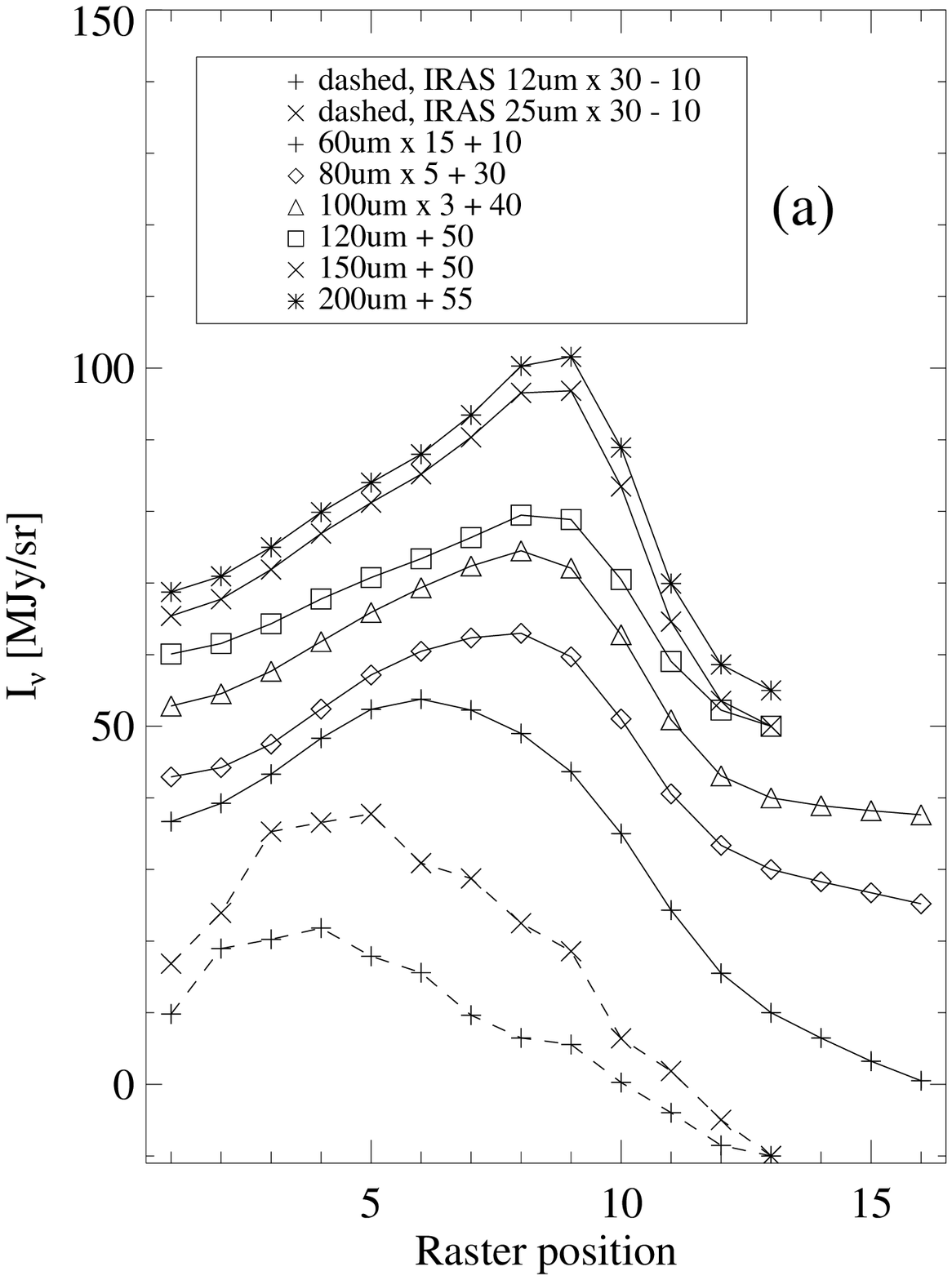}{11.5cm}{0}{63}{63}{-151}{-14}
\plotfiddle{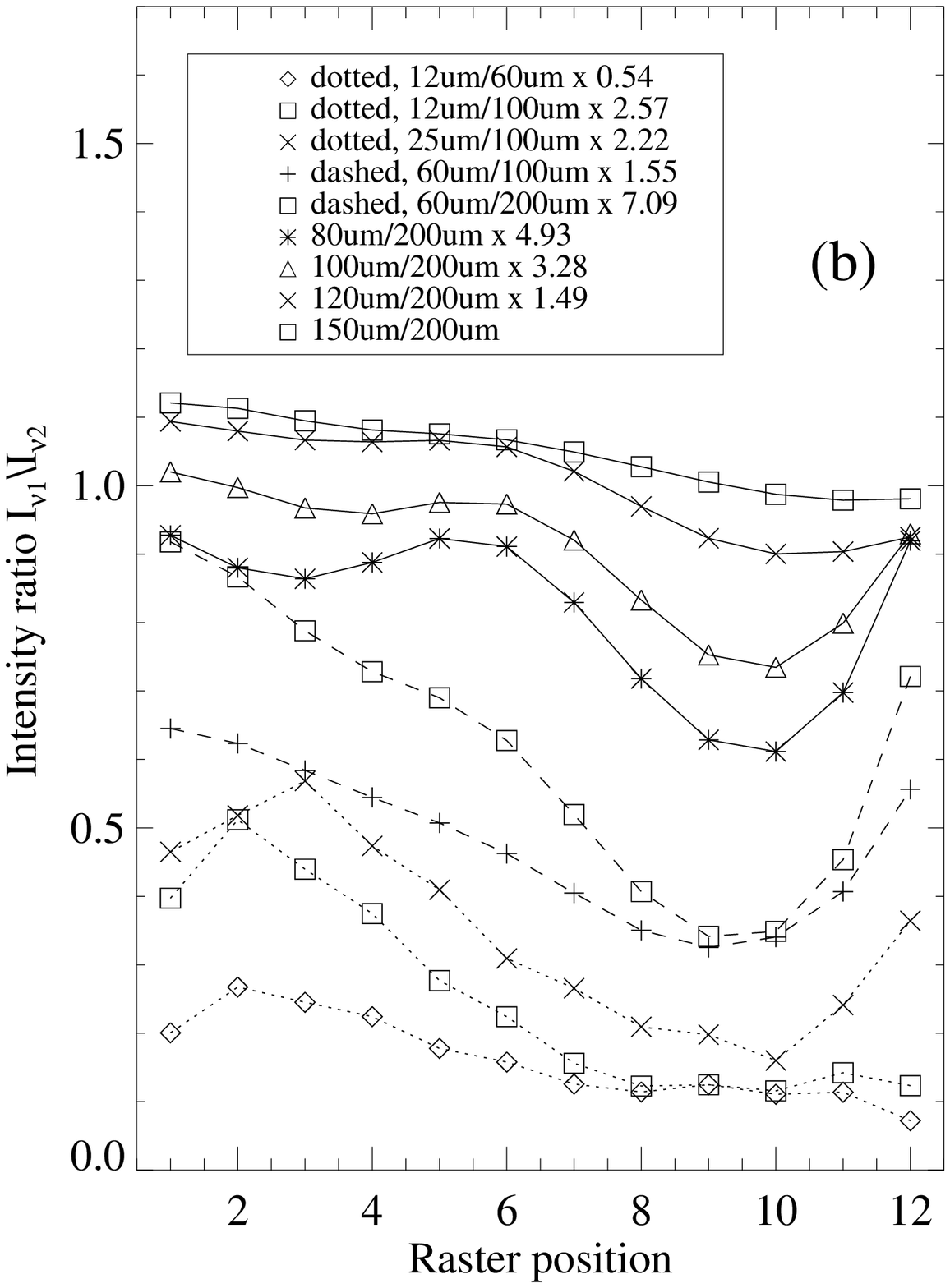}{11.5cm}{0}{63}{63}{-151}{-12}
\caption[]{(a) Intensities along the six ISO raster scans of L1780, plotted at the middle positions of the forth and back-going scans shown in Fig.~\ref{fig:multiplot}f (see Section 3.2). The 60$\mu$m, 80$\mu$m and 100$\mu$m scans have 16 positions, others 13. (b) Ratios of the raster scan intensities. In the plots, West is right, East is left.}
\label{fig:multiplot3}
\end{figure}

\begin{figure}
\centering 
\resizebox{\hsize}{!}{\includegraphics{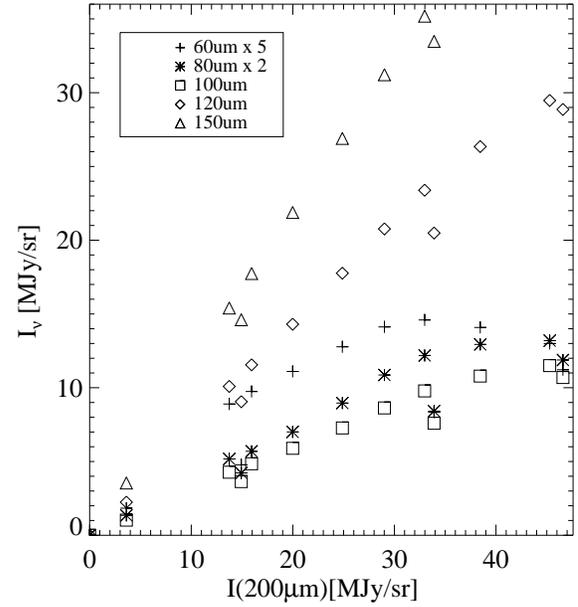}}
\caption{Intensities of the ISO raster scans vs. the 200$\mu$m raster scan intensity.} 
\label{fig:200stripevsothers} 
\end{figure} 

\begin{figure}
\centering 
\resizebox{\hsize}{!}{\includegraphics{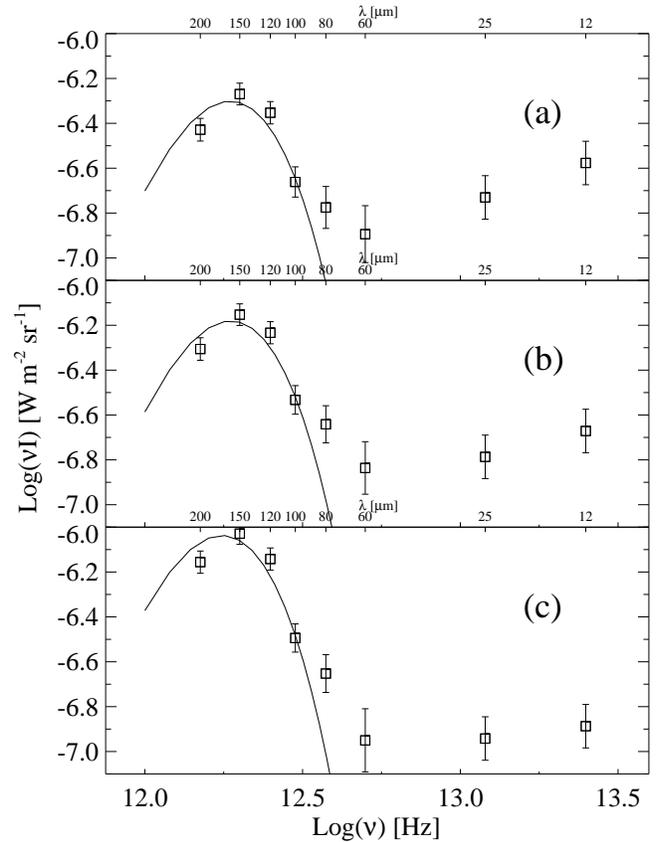}}  
\caption{Spectral energy distributions at the positions of the IRAS 12$\mu$m (a), IRAS 60$\mu$m (b) and ISO 100$\mu$m (c) maxima. The solid line shows a blackbody with $\lambda^{-2}$ emissivity law fitted to the 100$\mu$m, 120$\mu$m, 150$\mu$m and 200$\mu$m intensities. The error bars show the 1-$\sigma$ errors used in fitting (see the text).} 
\label{fig:SEDsatmaxima}
\end{figure}

\section{Results}

\subsection{General structure of the FIR emission from L1780}
The 100$\mu$m and 200$\mu$m surface brightness maps are shown in 
Fig.~\ref{fig:multiplot}a and Fig.~\ref{fig:multiplot}b. In the far-infrared, 
L1780 has a clearly visible brighter central core in the West, which
coincides with the region of maximum optical extinction. On the opposite
side, the cloud has an extended, tail-like structure, and an overall 
cometary shape.

The relation between the 100$\mu$m and the 200$\mu$m emission is shown 
in Fig.~\ref{fig:c200vsc100}. The 100$\mu$m emission is well 
correlated with the 200$\mu$m emission, and the relation is linear 
up to the highest surface brightness values, where there is a slight 
flattening trend. The solid lines in the figure show the 
surface brightness ratios at the indicated temperatures, assuming a modified
blackbody function with dust emissivity power law index $\beta=2$. 

As can be seen in Fig.~\ref{fig:multiplot}f,
the IRAS 60$\mu$m maximum is distinct from the 100$\mu$m\slash200$\mu$m 
maximum, and is situated East of it. 
The 25$\mu$m, the 12$\mu$m and the 6.7$\mu$m maxima are 
located still further to the East, reaching towards the SE edge 
of the cloud. The 100$\mu$m and 200$\mu$m maps do not show any structure
at the locations of these maxima.

The $^{13}$CO core of L1780 is in virial equilibrium (T\'oth et al. 1995).
However, no point sources, neither those from the IRAS Point Source Catalog
nor any new ones, are detectable in the ISO maps. Also, a J-H vs. H-K color
excess plot made using 2MASS data reveals no stars with color excess $\ge$1.0
mag in L1780. This leads to the conclusion that there is no (detectable) star
formation activity in L1780.

\subsection{Surface brightness ratios from ISO raster scans and IRAS observations}

In Fig.~\ref{fig:multiplot}f, the positions of the ISO raster 
scans, which go through the cloud from West to East and back, are shown.
Fig.~\ref{fig:multiplot3}a shows the intensities of the raster scans
averaged (using Gaussian beams with FWHM=4.5$\arcmin$) 
at the positions marked with circles in Fig.~\ref{fig:multiplot}f.

All the raster scans go through the high density core of L1780 and the 
regions of the maximum 12$\mu$m, 25$\mu$m and 60$\mu$m emission. The maxima of
the 120$\mu$m-150$\mu$m emission coincide with the maximum of the 200$\mu$m
emission, whereas the positions of the 60$\mu$m and the 80$\mu$m maxima are
shifted towards East. The shape of the 100$\mu$m brightness profile is between
the shapes of the 80$\mu$m and the 120$\mu$m, especially in the region of
maximum emission.

Fig.~\ref{fig:multiplot3}a also shows the values of the 
IRAS 12$\mu$m and 25$\mu$m observations calculated at the same 
positions as the raster scans. The maximum of the IRAS 25$\mu$m emission
is located between the 60$\mu$m maximum and the 12$\mu$m maximum 
on the Eastern side of L1780.

Fig.~\ref{fig:multiplot3}b shows the intensity ratios 12$\mu$m\slash60$\mu$m; 
$I_{\nu}$\slash100$\mu$m, where $I_{\nu}$ is 12$\mu$m, 25$\mu$m or 60$\mu$m; 
and $I_{\nu}$\slash200$\mu$m, where $I_{\nu}$ is 60$\mu$, 80$\mu$m, 100$\mu$m,
120$\mu$m or 150$\mu$m, plotted through the cloud from East to West (the
intensities have been calculated as in Fig.~\ref{fig:multiplot3}a). The cold
core of L1780 shows up as a minimum in all the curves. 
The ratios $I_{\nu}$\slash100$\mu$m and 60$\mu$m\slash200$\mu$m 
rise steeply towards the Eastern side of the
cloud, indicating an increased presence of the small-grain dust component. 
The ratio 12$\mu$m\slash60$\mu$m peaks at the maximum position of the
12$\mu$m emission in L1780, indicating the presence of PAHs in this location.
The ratios 80$\mu$m\slash200$\mu$m and 100$\mu$m\slash200$\mu$m show a
deep minimum at the core of 1780, while the ratios 120$\mu$m\slash200$\mu$m
and 150$\mu$m\slash200$\mu$m show only a slight drop at this location, the
150$\mu$m\slash200$\mu$m ratio being almost constant.

In Fig.~\ref{fig:200stripevsothers}, the 60-150$\mu$m surface brightness
values of the raster scans are plotted against the 200$\mu$m scan values.
All the wavelengths correlate well with the 200$\mu$m intensity.
However, with the exception of the 150$\mu$m vs. 200$\mu$m relation, all
curves clearly show signs of flattening close to the 200$\mu$m maximum.

\begin{table*}
\caption[]{The surface brightness values and corresponding errors used in Fig.~\ref{fig:SEDsatmaxima}. The three raster positions are given according to Fig.~\ref{fig:multiplot3}: the raster positions no. 4, 6 and 9 correspond to the 12$\mu$m, 60$\mu$m and 100$\mu$m emission maxima of L1780, respectively.}
\begin{center}
\begin{tabular}{rrrrrrr}
\hline\noalign{\smallskip}
 & \multicolumn{2}{c}{Raster position 4} & \multicolumn{2}{c}{Raster position 6} & \multicolumn{2}{c}{Raster position 9}\\
\hline\noalign{\smallskip}
$\lambda$ & $I(\nu)$[MJy\slash sr] & $\delta I(\nu)$[MJy\slash sr] & $I(\nu)$ & $\delta I(\nu)$ & $I(\nu)$ & $\delta I(\nu)$\\
\noalign{\smallskip}
\hline\noalign{\smallskip}
12$\mu$m & 1.06 & 0.21 & 0.85 & 0.17 & 0.52 & 0.10\\
25$\mu$m & 1.55 & 0.31 & 1.36 & 0.27 & 0.95 & 0.19\\
60$\mu$m & 2.56 & 0.65 & 2.92 & 0.69 & 2.24 & 0.62\\
80$\mu$m & 4.48 & 0.87 & 6.09 & 1.06 & 5.94 & 1.05\\
100$\mu$m & 7.27 & 1.04 & 9.78 & 1.34 & 10.69 & 1.44\\
120$\mu$m & 17.76 & 1.93 & 23.39 & 2.51 & 28.86 & 3.09\\ 
150$\mu$m & 26.90 & 2.84 & 35.19 & 3.69 & 46.84 & 4.90\\
200$\mu$m & 24.88 & 2.73 & 32.98 & 3.60 & 46.59 & 5.02\\
\noalign{\smallskip}
\hline
\end{tabular}
\end{center}
\label{tab:threespectra}
\end{table*}

\subsection{Dust temperature and the spectral energy distribution}
The temperature of dust has been derived, pixel by pixel, from the
100$\mu$m and 200$\mu$m ISO surface brightness maps by using a modified 
blackbody function
\begin{equation}
I(\lambda)\propto\lambda^{-\beta}B(\lambda,T_{\rm dust}),
\label{temper}
\end{equation}
where $\beta$ is the dust emissivity power law index, and 
$B(\lambda,T_{\rm dust})$ is the Planck function. We assume
$\beta=2$.
For the temperature calculation, the 100$\mu$m map was convolved to the 
resolution of the 200$\mu$m map. 
The surface brightness values were iteratively color
corrected, using the color correction coefficients 
based on previous temperature determination, until 
the differences between two consecutive iterations were 
below 0.1 K. The temperature map is shown in Fig.~\ref{fig:multiplot}e.
At the resolution of 1.5$\arcmin$, the minimum temperature is 14.9$\pm0.4$ K.

The temperature of L1780 has also been determined using the ISO raster scans.
Fig.~\ref{fig:SEDsatmaxima} shows the SEDs at three positions along the scan: 
Figs.~\ref{fig:SEDsatmaxima}a, \ref{fig:SEDsatmaxima}b and 
\ref{fig:SEDsatmaxima}c correspond to the locations of the 
IRAS 12$\mu$m, the IRAS 60$\mu$m and the ISO 100$\mu$m 
emission maxima, respectively (see Fig.~\ref{fig:multiplot}f 
for the locations of these maxima in L1780).
In addition to the ISO raster scan values at 60-200$\mu$m, 
the IRAS 12$\mu$m and 25$\mu$m values are plotted in the figure.
Modified blackbody curves with dust emissivity index $\beta =2$ have 
been fitted to the four longest wavelengths, resulting in
temperatures 15.0$\pm 0.4$ K, 15.0$\pm 0.4$ K 
and 14.1$\pm 0.4$ K, for Figs.~\ref{fig:SEDsatmaxima}a, 
\ref{fig:SEDsatmaxima}b and \ref{fig:SEDsatmaxima}c, respectively.
These temperatures are in agreement with the values obtained using 
the large 100$\mu$m and 200$\mu$m ISO maps.
In the fitting, the statistical errors (from PIA) plus 
filter-to-filter errors of 10\% have been used 
for the ISO C200 values (120$\mu$m, 150$\mu$m and 200$\mu$m). The error used 
for the 100$\mu$m values is the statistical error plus a 10\% error 
resulting from the uncertainty of the scaling of the 100$\mu$m to DIRBE.   
The average $\chi^2$ value of all the fits was slightly below 2.
In Fig.~\ref{fig:SEDsatmaxima} the 80$\mu$m are 
$\sim$2$\sigma$ above the fitted curves.
The ratio of the mid-infrared intensity (average of the
12$\mu$m and 25$\mu$m intensities) to the FIR intensity (maximum of the 
blackbody fit at the three longest wavelengths) is 0.045 for the IRAS 12$\mu$m 
maximum region (Fig.~\ref{fig:SEDsatmaxima}a), 0.029 for the IRAS 60$\mu$m maximum (Fig.~\ref{fig:SEDsatmaxima}b) 
and 0.013 for the ISO 100$\mu$m maximum (Fig.~\ref{fig:SEDsatmaxima}c).
All the surface brightnesses and errors used in Fig.~\ref{fig:SEDsatmaxima} are given in Table~\ref{tab:threespectra}. 


\begin{figure*}
\begin{minipage}{12cm}
\plotfiddle{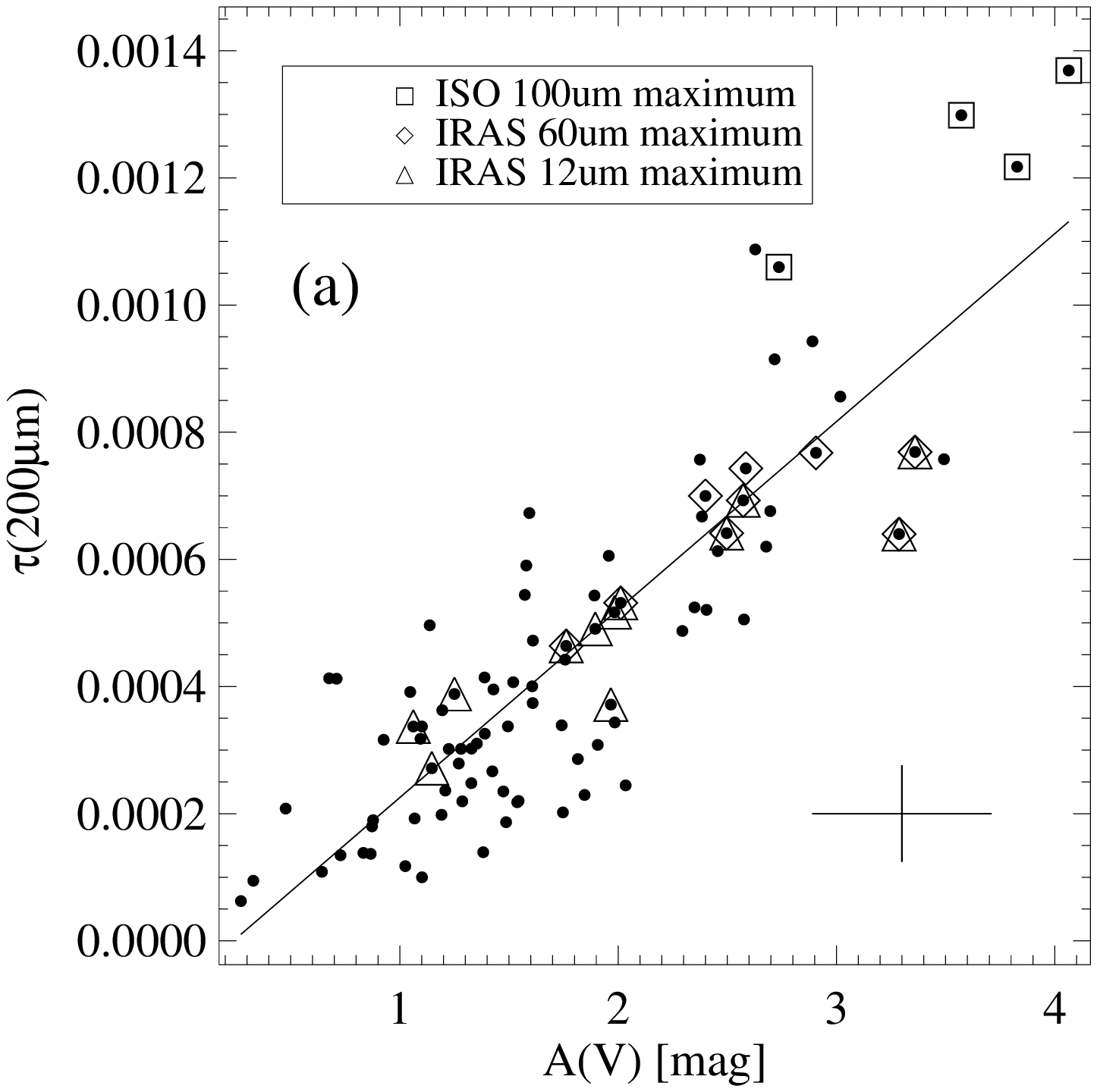}{7.7cm}{0}{60}{60}{-180}{-30}
\plotfiddle{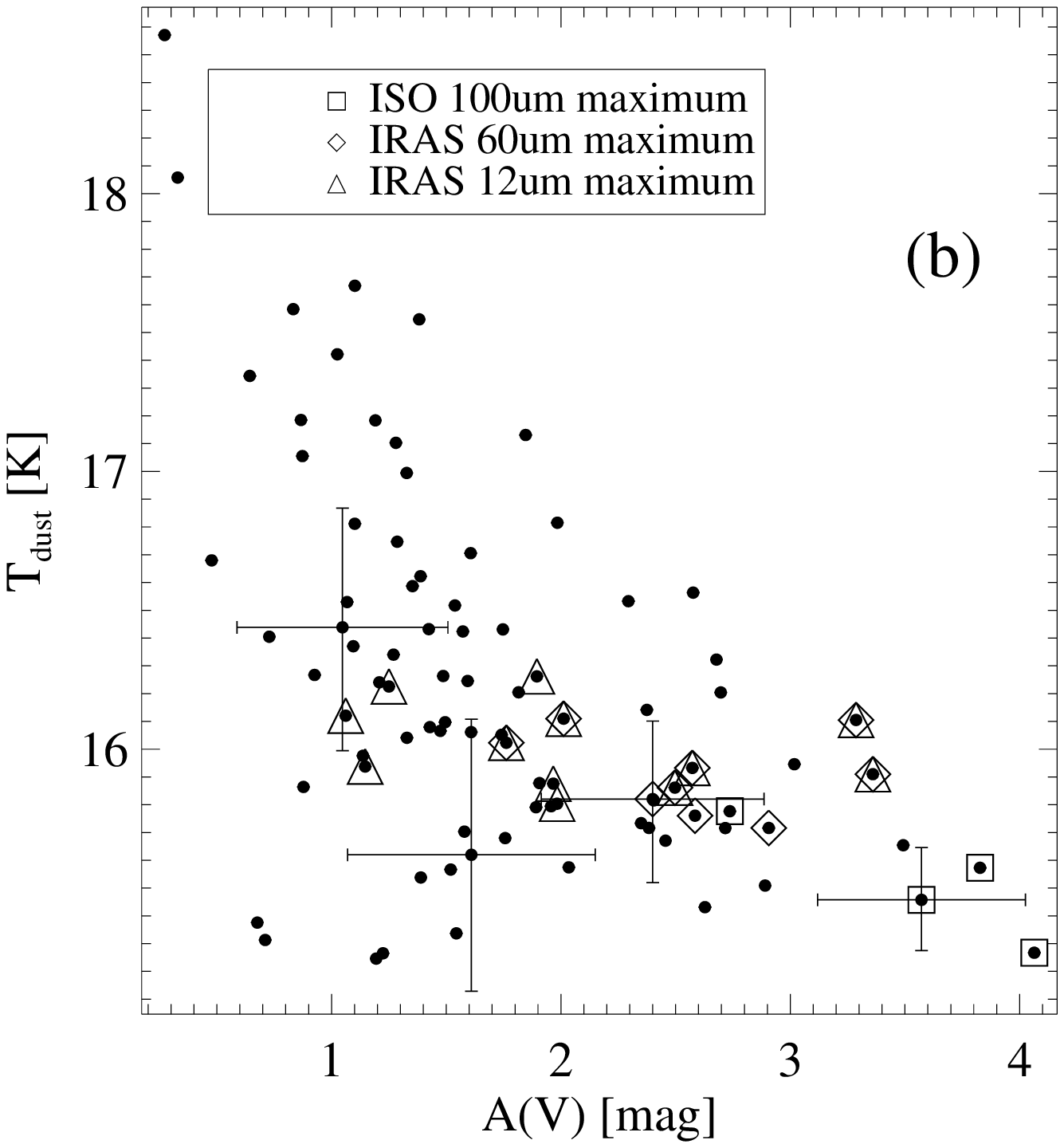}{0cm}{0}{60}{60}{+73}{-8}
\plotfiddle{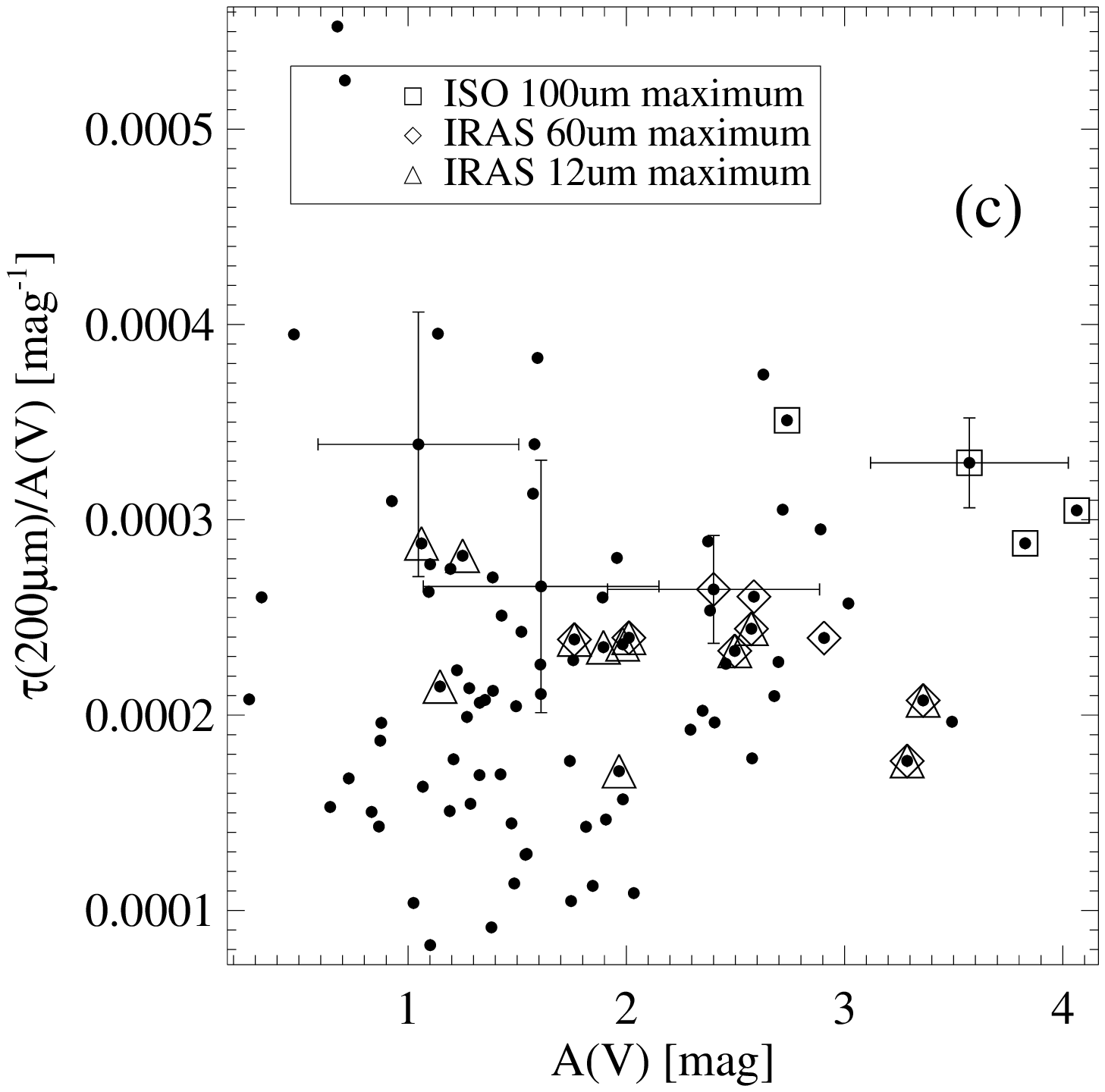}{7.8cm}{0}{60}{60}{-180}{-30}
\plotfiddle{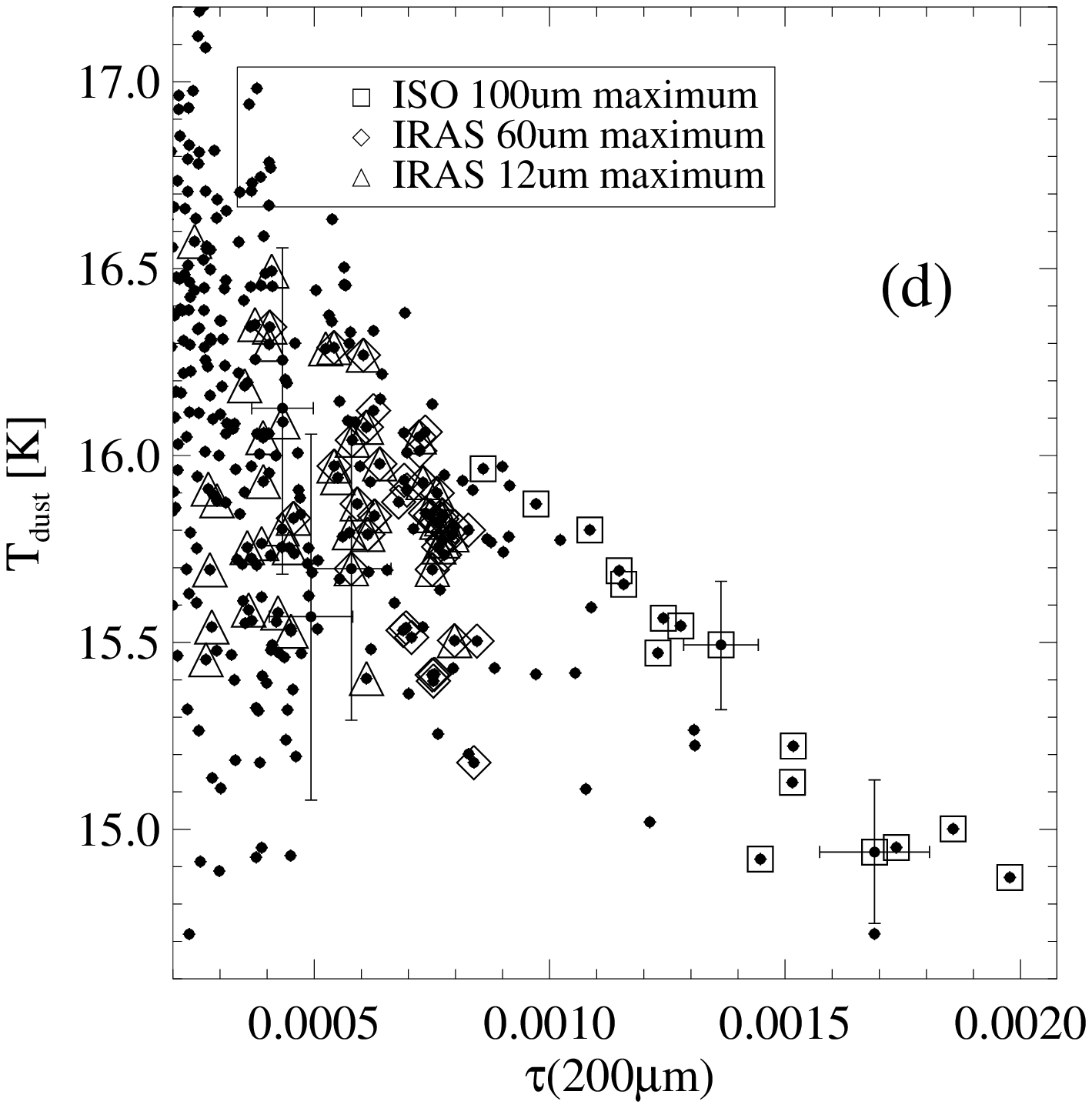}{0cm}{0}{60}{60}{+73}{-8}
\end{minipage}
\caption[]{(a) The 200$\mu$m optical depth versus the visual extinction $A_V$, at 3.0$\arcmin$ resolution. For the linear fit, see Section 3.4. (b) Temperature versus the visual extinction. (c) $\tau({\rm 200\mu{\rm m})}\slash A_V$ versus $A_V$. (d) Temperature versus the 200$\mu$m optical depth, at 1.5$\arcmin$ resolution. For (a), (b), (c) and (d), squares, diamonds and triangles mark the points corresponding to the ISO 100$\mu$m, IRAS 60$\mu$m and IRAS 12$\mu$m surface brightness maxima in the sky, respectively. The error bars show typical errors.}
\label{fig:multiplot2}
\end{figure*}

\subsection{FIR-optical depth and cloud mass}
The optical depth at 200$\mu$m has been calculated from the equation
\begin{equation} 
\tau(200\mu{\rm m})=\frac{I(200\mu{\rm m})}{B(200\mu{\rm m},T_{\rm dust})}, 
\label{optdepth} 
\end{equation} 
which is valid for optically thin emission and where $I$ is the observed 
intensity and $B(T_{\rm dust})$ the intensity of a blackbody at the
temperature $T_{\rm dust}$. The optical depth map is 
shown in Fig.~\ref{fig:multiplot}d. The maximum value of the 
200$\mu$m optical depth is 2.0$\pm 0.3\times 10^{-3}$ 
at the resolution of 1.5$\arcmin$.

To calculate the mass of the cloud we first derived the ratio between the FIR
optical depth and the hydrogen column density, i.e. the average
absorption cross section per H-nucleus
\begin{equation}
\sigma^H(\lambda)=\frac{\tau(\lambda)}{N(H)}.
\label{sigmaH} 
\end{equation}
The NIR extinction can be used to estimate the total hydrogen column density,
$N(H+H_2)$ in L1780. We adopt the value 
$N(H+H_2)/E(B-V)=5.8\times10^{21}{\rm cm}^{-2}{\rm mag}^{-1}$, valid for
diffuse clouds (Bohlin, Savage \& Drake 1978), together with $A_V/E(B-V)=3.1$
to obtain $N(H+H_2)/A_V=1.87\times10^{21}{\rm cm}^{-2}{\rm mag}^{-1}$.
The relation between $A_V$ and $\tau(200\mu{\rm m})$ is shown in 
Fig.~\ref{fig:multiplot2}a.
A linear fit, 
$\tau(200\mu{\rm m})=3.0\times 10^{-4}\times A_V-6.4\times 10^{-5}$,
also shown in the figure, gives 
$\sigma^H(200\mu{\rm m})= 1.4\times 10^{-25}{\rm cm}^2$ per H-atom
for the absorption cross section, which is in agreement with values
for diffuse ISM by Dwek et al. (1997) 
($1.4\times 10^{-25}{\rm cm}^2$ per H-atom).
The relation in Fig.~\ref{fig:multiplot2}a deviates from 
linearity in the region of highest density in L1780; see Section 4.3
for discussion.

The total mass (gas plus dust) of L1780 has been calculated using the 
equation
\begin{equation}
M_{\rm FIR}=\frac{\tau(200\mu{\rm m})}{\sigma^H(200\mu{\rm m})}D^2m_{\rm H}\mu ,
\end{equation}
where $\tau_{200\mu{\rm m}}$ is the 200$\mu$m optical depth, 
$\sigma^H(200\mu{\rm m})$ is the absorption cross section 
per H-nucleus for which we have used the value derived above, 
$D$ is the distance,
$m_{\rm H}$ is the hydrogen mass, and $\mu$ is the mean molecular weight. 
The mass has been derived 
by summing up the pixels of the $\tau(200\mu{\rm m})$ map. 
The mass thus obtained for L1780 is $18 M_{\sun}$.

\begin{figure}
\centering  
\resizebox{\hsize}{!}{\includegraphics{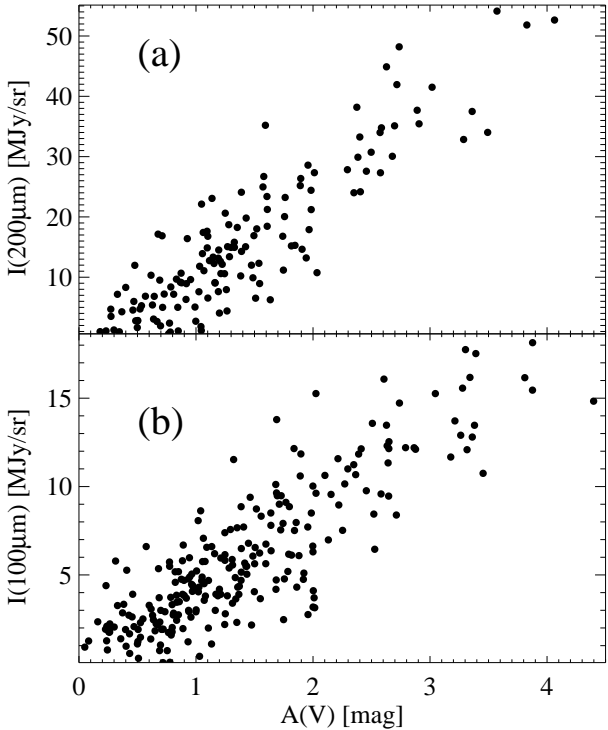}}
\caption{(a) ISO 200$\mu$m and (b) 100$\mu$m surface brightness values versus $A_V$.}
\label{fig:I100I200vsAv}
\end{figure}

\subsection{Comparison of FIR and extinction data}
At large scale, the far-infrared emission in L1780 correlates well with 
the optical extinction.
The 100$\mu$m, 200$\mu$m and $A_V$ maps in 
Figs.~\ref{fig:multiplot}a, \ref{fig:multiplot}b 
and \ref{fig:multiplot}C show that the 100$\mu$m and the 200$\mu$m 
emission maxima coincide with the location of the highest visual extinction. 
The relations between the 100$\mu$m and the 200$\mu$m emission and
the visual extinction are linear (Fig.~\ref{fig:I100I200vsAv}).

In the $A_V$ map, there seems to be another region of higher visual 
extinction, an arc-like structure, on the Eastern side of the 100$\mu$m core. 
If real, this feature, together with the cometary shape of L1780, 
could result from the propagation of a shockwave from the SW direction through 
the cloud as has been suggested by T\'oth et al. (1995). In fact,
the arc-like feature could also be a part 
of a ring-shaped structure, in which case 
an interesting analog of a possible dense-core formation would be found in 
the Globule 2 of the Coalsack cloud (Lada et al. 2004).
However, the extinction values 
may be biased because of variations in the stellar density. This
hampers the detection of morphological features in the $A_V$ map at this 
resolution, which is close to the detection limit 
($\Delta A_V\approx 2\sigma$ in the contours in Fig.~\ref{fig:multiplot}c).

\begin{figure}
\centering 
\resizebox{\hsize}{!}{\includegraphics{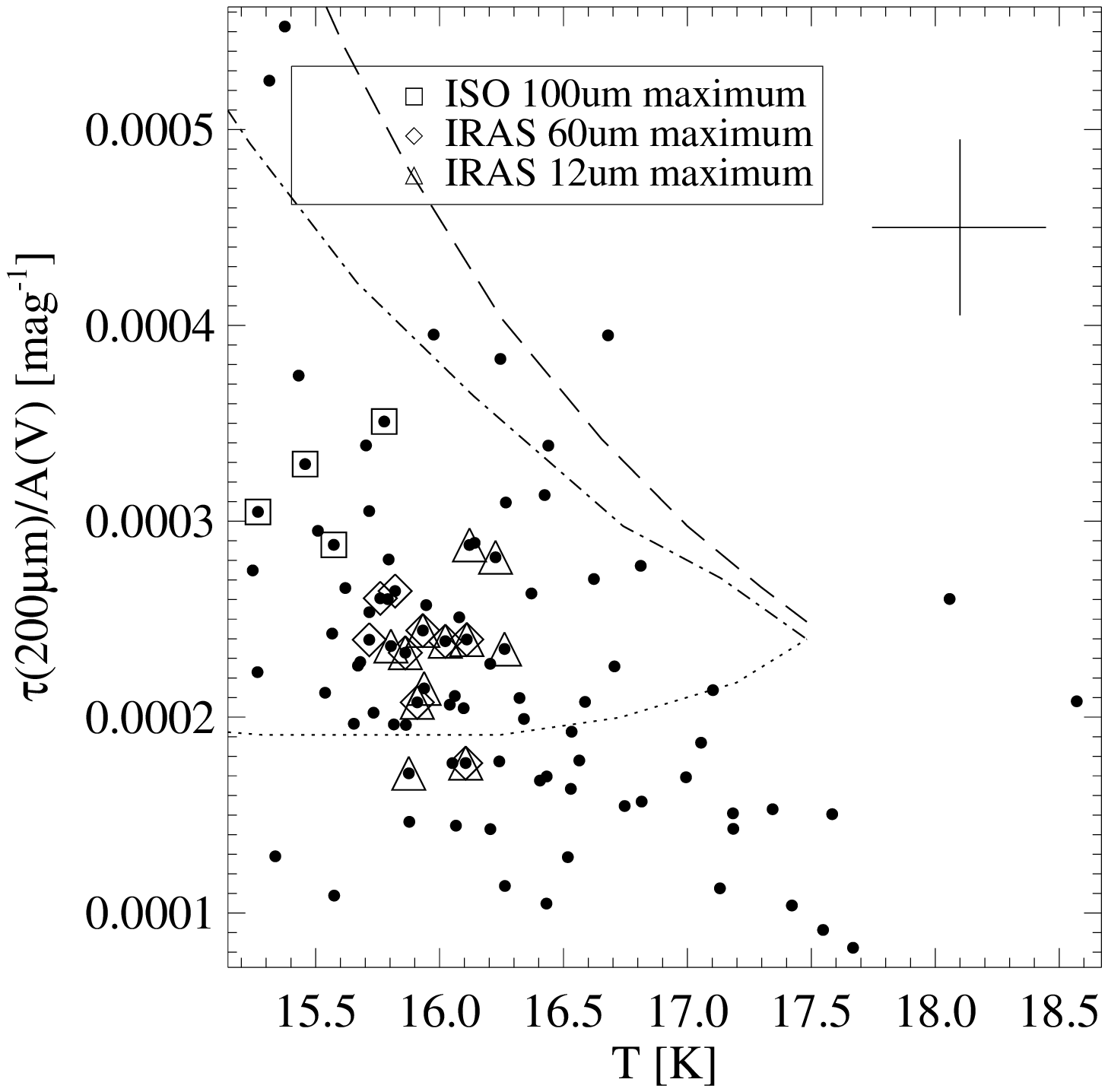}}  
\caption{$\tau({\rm 200\mu{\rm m})}\slash A_V$ versus T. The error bars plotted in the upper right corner show the magnitudes of typical errors. The curves show the emissivity increment to eight-fold (dashed line), four-fold (dash-dot line) and no increment (dotted line), when the relative amount of the cold component increases from 0.2 to 1 (from right to left), according to the model by del Burgo et al. (2003).}
\label{fig:tauperAvvsT}
\end{figure}

Figs.~\ref{fig:multiplot2} and \ref{fig:tauperAvvsT} 
show the relations between $A_V$, $\tau(200\mu$m) and $T_{\rm dust}$. 
The values from the locations of the 
IRAS 12$\mu$m, IRAS 60$\mu$m and ISO 100$\mu$m maxima are marked with 
triangles, diamonds and squares, respectively. The sizes of
sky areas of the maxima have been chosen according to the highest-level 
contours visible in Fig.~\ref{fig:multiplot}f, and are 
$12.5\arcmin\times 3.0\arcmin$, $11.3\arcmin\times 4.0\arcmin$, and 
$6.3\arcmin\times 4.0\arcmin$ 
for the 12$\mu$m, 60$\mu$m and 100$\mu$m maximum, respectively. 
The error bars in the figures show typical errors. The errors of the 
temperature and the 200$\mu$m optical depth get larger for 
the points near to the borders of the maps, since
near the borders of the ISO FIR maps the background-subtracted surface
brigthness values, used for calculating $T_{\rm dust}$ and $\tau(200\mu$m),
are more uncertain. For the plots in Figs.~\ref{fig:multiplot2} 
and \ref{fig:tauperAvvsT}, most of the near-border points have been removed. 

Fig.~\ref{fig:multiplot2}d shows the temperature decrease in the densest 
part of the cloud: as the optical depth increases, the temperature
decreases from $\sim$15.8 K in 
the regions of the 12$\mu$m emission maximum (PAHs) and the 
60$\mu$m maximum (VSGs) to $\sim$14.9 K in 
the 100$\mu$m maximum region (big grains). 
In Fig.~\ref{fig:multiplot2}b, similar dependence is shown 
for the temperature as the function of the visual extinction.
Fig.~\ref{fig:multiplot2}a shows increased emissivity
in the regions of high $A_V$ in L1780 (see Section 4.3 for discussion).

\begin{figure}
\centering  
\resizebox{\hsize}{!}{\includegraphics{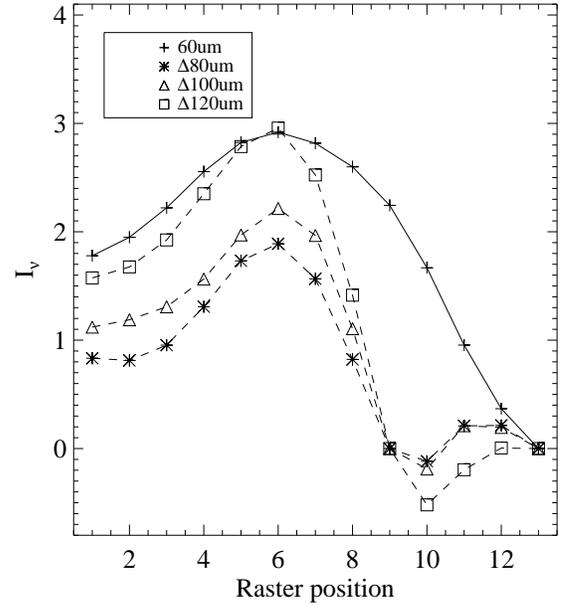}}
\caption{Comparison between the 60$\mu$m ISO raster scan and the 80-120$\mu$m raster scans from which the 200$\mu$m emission (the large-grain contribution) has been subtracted.}
\label{fig:smallgraincontrib}
\end{figure}

\begin{table*}
\caption[]{The average and maximum values of the infrared colors along the ISO raster scans in L1780. The solar neighborhood (SN) values are from Boulanger et al. (1990).}
\begin{center}
\begin{tabular}{rrrrr}
\hline\noalign{\smallskip}
Color & $\mathrm{L1780_{AVG}}$ & $\mathrm{L1780_{MAX}}$ &  $\mathrm{L1780_{MIN}}$ & SN\\
\noalign{\smallskip}
\hline\noalign{\smallskip}
12\slash 100 & 0.10 & 0.20 & 0.05 & 0.042\\
25\slash 100 & 0.16 & 0.26 & 0.07 & 0.054\\
60\slash 100 & 0.31 & 0.42 & 0.21 & 0.21\\
12\slash 25 & 0.59 & 0.85 & 0.29 & \\
\noalign{\smallskip}
\hline
\end{tabular}
\end{center}
\label{tab:colors}
\end{table*}

\section{Discussion}
\subsection{Different dust populations from FIR observations}
The Southern side of L1780 is facing both the Galactic plane in the 
South and the Upper Scorpius (USco) 
association of OB stars in the SW direction, and is, therefore,
subject to more intense ISRF.
However, the variations in the emission of L1780 at different wavelengths 
are more prominent in the E-W direction than on the S-N axis.
Fig.~\ref{fig:multiplot}f shows that the 12$\mu$m emission 
maximum (indicating the
presence of PAHs) is located on the SE side, in the ``tail'' of L1780. 
The 60$\mu$m (VSGs) emission maximum is located to the West from 
the 12$\mu$m maximum, and finally, the 100$\mu$m maximum (big grain emission) 
is on the Western side of the cloud.

In Fig.~\ref{fig:multiplot3}a, 
the 60$\mu$m emission curve clearly rises towards
East and has its maximum on the Eastern side of the 200$\mu$m maximum.
Fig.~\ref{fig:multiplot3}b shows that the 60$\mu$m\slash100$\mu$m ratio 
rises rather steeply from its minimum towards the East side of L1780. In 
Table~\ref{tab:colors}, it can be seen that the minimum value of
this color ratio (reached in the cold core of L1780) is the same as the solar 
neighborhood (SN) value by Boulanger et al. (1990).
Laureijs, Clark \& Prusti (1991), who made IRAS
observations on the L134 complex, indicate in their study of L1780
that the 60$\mu$m emission diminishes in a narrow transition 
layer around the 100$\mu$m maximum.
Our data do not show any abrupt change, but the decrease 
in the 60$\mu$m\slash100$\mu$m ratio is clear: the ratio is halved 
in the cold core region relative to the Eastern edge of the cloud.
It is not likely that this reduction would be due to attenuation effects 
alone: in the model by Bernard et al. (1992) the effect of radiation 
attenuation on the ratio 60um\slash100$\mu$m as Av increases from 1 to 4 mag 
(between different 1D models) 
is only half of the decrease of the ratio observed in L1780.

Laureijs, Clark \& Prusti (1991) attribute the change in 60um\slash100$\mu$m 
to the formation of icy mantles on grains. 
Stepnik et al. (2003) shows that the depletion of VSGs 
through grain-grain coagulation can reduce the 60um\slash100$\mu$m ratio, and,
in addition, cause an increased FIR emissivity of grains, 
which is also observed in L1780.
Fig.~\ref{fig:multiplot2}c shows the increased values of $\tau_{200}/A_V$ 
in the region of the 100$\mu$m emission maximum 
(marked with squares in Figs.~\ref{fig:multiplot2}b 
and \ref{fig:multiplot2}c). On the other hand, 
if taken to be traced by the $\tau_{200}/A_V$ ratio, 
no difference in the overall size distribution 
of the large grains between the 60$\mu$m and 12$\mu$m maxima is visible in
Fig.~\ref{fig:multiplot2}c, although the 60$\mu$m maximum 
is located deeper in the cloud.

Figs.~\ref{fig:multiplot3}a and \ref{fig:multiplot3}b show that
the emission from VSGs and PAHs is concentrated on the Eastern side of L1780,
and that the PAH emission clearly is distinct from the VSG emission.
Also the MIR\slash FIR ratios from Fig.~\ref{fig:SEDsatmaxima} indicate 
increase of the 12$\mu$m-60$\mu$m emission toward the East.
For the 12$\mu$m\slash100$\mu$m ratio, comparison with the results 
of the model of Bernard et al. (1992) suggests that radiative transfer 
effects account for less than one third of the 
decrease of the ratio in the L1780 core.
Table~\ref{tab:colors} shows that the average colors 
12$\mu$m\slash100$\mu$m and 25$\mu$m\slash100$\mu$m are more than 
twice the SN values. 
At the scan positions of the 12$\mu$m and 25$\mu$m maximum emission,
the values of 12$\mu$m\slash100$\mu$m and 25$\mu$m\slash100$\mu$m 
are 0.15 and 0.18, corresponding to 3.6 and 3.3 times the SN value. 
The maximum 12$\mu$m\slash100$\mu$m and 
25$\mu$m\slash100$\mu$m ratios observed 
in L1780 are both 4.8 times the SN value. 
These values can be compared with
those of two clouds from Chameleon and $\rho$-Ophiuchi, which are 
5.5 and 3.3 times the SN value for PAHs 
(using the 12$\mu$m\slash100$\mu$m ratio), respectively, 
and 2.3 and 2.3 times the SN value 
for VSGs (using the 25$\mu$m\slash100$\mu$m ratio), respectively, 
(Bernard, Boulanger \& Puget 1993). 
The color ratios observed in L1780 indicate that there is 
an overabundance of PAHs and VSGs on the Eastern side of the cloud. 
Further support to our observations on PAH abundance is provided 
by the strong decrease of the 6.5\slash100$\mu$m ratio in the FIR core of L1780 
observed by Miville-Desch\^enes (2002) using ISOCAM and IRAS observations.

We conclude that, although attenuation effects may be involved, 
they are not sufficient to explain the observed color ratios, which indicate 
true variations in the abundances of dust grains from different populations. 
To estimate the effect of radiation attenuation on the observed
color variations in L1780, model calculations are needed 
(Ridderstad et al., in preparation).

\subsection{VSG contribution at the wavelengths above 80$\mu$m}

Radiative transfer modelling performed using the current ISM models 
(including PAHs, VSGs and big grains) indicates that the emission from 
VSGs is still significant above 60$\mu$m, and contributes from 15\% to 23\%
of the total emission at 100 $\mu$m when $A_V$ ranges from 1 to 4 mag 
(D\'esert, Boulanger \& Puget 1990).
Laureijs et al. (1996) made ISOPHOT observations 
at 60$\mu$m, 90$\mu$m, 135$\mu$m and 200$\mu$m
on a small cloud in Chameleon, 
which, like L1780, has a reduced 60$\mu$m\slash100$\mu$m
ratio in the centre, and indicated that the 90$\mu$m emission has an excess
contribution from dust grains emitting at 60$\mu$m. 

In Fig.~\ref{fig:smallgraincontrib}, the 200$\mu$m raster scan
has been assumed to contain emission solely from the big-grain dust component,
which then has been subtracted from the 80$\mu$m, 100$\mu$m and 120$\mu$m 
raster scans after scaling the latter using a ratio 
$I^C_{\nu}$\slash$I^C_{200\mu{\rm m}}$ determined at the cloud 
center, i.e. at the 200$\mu$m maximum position 
(position 9 in Figs.~\ref{fig:multiplot3}a):
\begin{equation}
\Delta I_{\nu} = I_{\nu} - \frac{I^C_{\nu}}{I^C_{200\mu{\rm m}}}\times I_{200\mu{\rm m}} \quad \mbox{MJy\slash sr}.
\end{equation}
The remaining part, $\Delta I_{\nu}$, represents excess that
can be attributed to small grain emission, assuming that no temperature 
change of big grains 
resulting from the increase of $A_V$ is taken into account. 
In reality, the $\Delta I_{\nu}$ always
contain also the effect from radiation attenuation; thus, 
Fig.~\ref{fig:smallgraincontrib} gives an upper limit to 
the VSG contribution. 
Fig.~\ref{fig:smallgraincontrib} shows that 
the 200$\mu$m-subtracted intensities peak at the maximum position of 
the 60$\mu$m emission. Our results
indicate that, at the maximum position of the 60$\mu$m emission, 
at most 12\% of the 120$\mu$m emission, 
22\% of the 100$\mu$m emission and 28\% 
of the 80$\mu$m emission is from VSGs. When it is taken into account that 
the ISO filter bandpasses overlap, this is consistent with the current 
ISM dust model (see D\'esert, Boulanger \& Puget 1990).

\subsection{Increased FIR emissivity in the dense core of L1780}

Many recent studies indicate a change in the emissivity 
of interstellar grains in dense molecular regions (Cambr\'esy et al. 2001; 
del Burgo et al. 2003; Dupac et al. 2003; Kramer et al. 2003; 
Stepnik et al. 2003; Cambr\'esy, Jarrett \& Beichman 2005). 
Also in our study, 
Fig.~\ref{fig:multiplot2}a shows that in the cold core of L1780 
(the region of the FIR emission maximum and minimum temperature 
$T_{\rm dust}$) 
above $A_{\rm V}\approx 3.5$, the relation 
between $\tau(200\mu$m) and $A_V$ deviates from linearity,
indicating an increased emissivity of the big grains.
This increase in the ratio $\tau(200\mu$m)\slash$A_V$ in the 100$\mu$m 
emission maximum area is also visible in Fig.~\ref{fig:multiplot2}c. 
The emissivity observed in the cold core of L1780 
in Fig.~\ref{fig:multiplot2}a is $\sim$1.5 times the values observed
elsewhere in the cloud. This value is in agreement with the results of
Cambr\'esy, Jarrett \& Beichman (2005), who considered the whole galactic
anticenter hemisphere and found that, for regions with $A_V>1$ mag, 
the ratio of FIR optical depth to NIR extinction is 
$A_V$(FIR)\slash$A_V$(gal)=1.31$\pm0.06$.  

Del Burgo et al. (2003) found FIR emissivity changes in their
60-200$\mu$m observations of eight regions mostly belonging 
to quiescent high latitude clouds with $A_V\sim 1-6$ mag. Since L1780
is also a translucent cloud, their results are of particular interest here.
Del Burgo et al. (2003) presented a
model with two big-grain dust components: a warm component at T=17.5 K 
and a cold component at T=13.5 K. A coefficient $\epsilon$ gives the factor
by which the emissivity of the cold component is increased.
In Fig.~\ref{fig:tauperAvvsT}, the curves showing the emissivity change 
according to the model by del Burgo et al. (2003) are shown for comparison
with our data. Comparison with the results of del Burgo et al. (2003) indicates
that the emissivity in the core of L1780 is about twice of that for the 
diffuse ISM, which is in agreement with the value derived from the
relation in Fig.~\ref{fig:multiplot2}a. 
Although the uncertainty in Fig.~\ref{fig:tauperAvvsT} 
increases with increasing temperature,
it is clear from the average errors (shown by error bars in the figure)
that there is an increasing trend towards high $\tau_{200}/A_V$
values. It must be noted that in this kind of plot,
the dependance of the optical depth on the dust temperature (since the 
former is calculated using the latter) can cause false correlation 
similar to the supposed emissivity increase. However, the errors
in the plot in Fig.~\ref{fig:tauperAvvsT} diminish towards high optical 
depth and low temperature. 
We also have the advantage that the measurements are from a single cloud,
which eliminates the effects resulting from calibration errors between
separately calibrated maps.

The ratio $\tau_{200}/A_V$ increases with grain size and porosity 
(Cambr\'esy et al. 2001), and
the change in FIR\slash submm emissivity has been attributed to 
the coagulation of grains into larger, fluffy particles in 
dense, cold molecular regions (Cambr\'esy et al. 2001, Stepnik et al. 2003). 
It is interesting that an independent indicator, the decrease of the 
60$\mu$m\slash100$\mu$m ratio towards the dense core of
many clouds, including L1780, has also been attributed to grain 
coagulation (Bernard et al. 1999, Stepnik et al. 2003). 
While the emissivity change in L1780
is seen above $\sim$3.5 mag, Stepnik et al. (2003) derived 
the threshold value $A_V=2.1\pm0.5$ for the possibly 
coagulation-induced changes (reduced 60$\mu$m\slash100$\mu$m ratio 
and increased emissivity) in the properties of the grains to occur.  
They noted that these changes in the properties of dust grains
may be part of a general transformation process that dust 
undergoes in the dense ISM. Moreover, Cambr\'esy, Jarrett \& Beichman (2005)
indicate that large, fluffy grains may be common all over the galaxy
wherever $A_V\geq 1$.

\section{Conclusions}
The analysis of the ISO 100$\mu$m and 200$\mu$m maps and 
60$\mu$m, 80$\mu$m, 100$\mu$m, 120$\mu$m, 150$\mu$m and 200$\mu$m 
raster scans of L1780, combined with the IRAS 12$\mu$m and  25$\mu$m
data and the 
visual extinction map (based on NIR $J$, $H$ and $K_{\rm S}$ band color 
excess data), leads to the following conclusions: 

   \begin{itemize}
      \item[-] The 100$\mu$m emission is well correlated with the 200$\mu$m 
emission throughout the cloud. The cloud has a single core, 
revealed by the 100$\mu$m and 200$\mu$m emission, 
the FIR optical depth, and the visual extinction. 
      \item[-] The spatial distributions of the 12$\mu$m, 25$\mu$m and
60$\mu$m emission differ significantly 
from the emission at longer wavelengths. This indicates
the presence of separate dust components with different 
physical properties and spatial distributions. 
      \item[-] The maximum values of the color ratios 12$\mu$m\slash100$\mu$m, 
25$\mu$m\slash100$\mu$m and 60$\mu$m\slash100$\mu$m, which 
are 4.8, 4.8 and 2.0 times the solar neighborhood values, indicate 
an overabundance of PAHs and VSGs in L1780. 
      \item[-] The $A_V$ map gives a maximum visual extinction
of 4 mag at 3.0$\arcmin$ resolution. The visual extinction is well correlated 
with the FIR emission at large scale.
      \item[-] The cold core of L1780 has the minimum temperature of 
14.9$\pm 0.4$ K and the maximum 200$\mu$m optical depth 
of 2.0$\pm 0.3\times 10^{-3}$ at 1.5$\arcmin$ resolution.
      \item[-] The value of the absorption cross section per 
H-atom at 200$\mu$m has been estimated to be 
$\sigma^H(200\mu{\rm m})=1.4\times 10^{-25}\mathrm{cm}^2$ per hydrogen 
atom, which is in good agreement with values obtained for diffuse ISM in 
other studies. The mass obtained for L1780 is 18$M_{\sun}$.
      \item[-] The relation between $A_V$ and $\tau(200\mu{\rm m})$ shows 
nonlinearity at high $A_V$ values, indicating that 
in the cloud core the far-infrared dust emissivity has increased by a factor
of $\sim$1.5.
     \item[-] The comparison of the 80$\mu$m and the 100$\mu$m emission with 
the emission at longer wavelengths indicate that not only the
80$\mu$m, but also the 100$\mu$m and even the 120$\mu$m emission
may contain some small-grain contribution.
   \end{itemize}

\begin{acknowledgements}
The work of M.R. has been supported by the Magnus Ehrnrooth 
Foundation and the Finnish Graduate School in Astronomy and Space Physics, 
which is gratefully acknowledged. The work of M.J. and K.L. has been
supported by the Finnish Academy through grants Nos. 17854 and 176071. 
We thank the referee, Jean-Philippe Bernard, for very helpful comments.\\
This publication makes use of data products from the Two Micron All 
Sky Survey, which is a joint project of the University of Massachusetts 
and the Infrared Processing and Analysis Center/California Institute of 
Technology, funded by the National Aeronautics and Space Administration 
and the National Science Foundation. Also, this research has made use of 
NASA\slash IPAC Infrared Science Archive.
\end{acknowledgements}

\end{document}